\documentclass[10pt]{article}
\usepackage{verbatim}

\usepackage{graphics}
\usepackage{graphicx}
\usepackage{color}

\usepackage{miller}
\usepackage{authblk}
\newcommand{\half}{\ensuremath{^1\!/\!_2}}

\title{An empirical potential for simulating hydrogen isotope retention in highly irradiated tungsten}

\author[1]{Daniel R. Mason \thanks {daniel.mason@ukaea.uk} }
\author[1]{Duc Nguyen-Manh}
\author[2]{Victor W. Lindblad}
\author[2]{Fredric G. Granberg}
\author[1]{Mikhail Yu. Lavrentiev}
\affil[1]{UK Atomic Energy Authority, Culham Centre for Fusion Energy, Oxfordshire OX14 3DB, United Kingdom}
\affil[2]{University of Helsinki Department of Physics, P.O.Box 43, FIN-00014, Finland}

\date{May 2023}

\begin{document}
\maketitle

\begin{abstract}
We describe the parameterization of a tungsten-hydrogen empirical potential designed for use with large-scale molecular dynamics simulations of highly irradiated tungsten containing hydrogen isotope atoms, and report test results.
Particular attention has been paid to getting good elastic properties, including the relaxation volumes of small defect clusters, and to the interaction energy between hydrogen isotopes and typical irradiation-induced defects in tungsten.
We conclude that the energy ordering of defects changes with the ratio of H atoms to point defects, indicating that this potential is suitable for exploring mechanisms of trap mutation, including vacancy loop to plate-like void transformations.
\end{abstract}

\section{Introduction}

For viable commercial D-T nuclear fusion, it is essential that the breeding of tritium at least balances the tritium which is retained after having penetrated the components.
One of the critical reactor components requiring study in this regard is the tungsten divertor, which is anticipated to undergo very high neutron flux~\cite{You2016,Gilbert_FST2014} and so develop a highly damaged microstructure, especially near coolant pipes where the temperature is kept below the onset of stage III recovery ( vacancy migration )~\cite{Keys_PhysRev_1968,Hirai_NME2016,Li_NME2018,You_JNM2021}. 
It is well established that irradiation-induced lattice defects within tungsten act as strong traps for hydrogen isotopes~\cite{Ogorodnikova_JNM2015,Ogorodnikova_JAP2015,DeBacker_PhysScr2017,DeBacker_NucFus2018,Wang_JNM2021,Hollingsworth_NF2019}.
While the modelling of binding energies of hydrogen to prototypical small defects is properly undertaken by density functional calculations, there is an emerging awareness that in the high irradiation dose limit the complex defect microstructure both generates and is responsive to stress~\cite{Mason_PRL2020}. 
This in turn can influence the evolution of defects~\cite{Kato_NucFus2015,SchwarzSelinger_NME2018} and lead to defect stabilization~\cite{Liu_AIPA2013,Qin_JNM2015,Zayachuk_NucFus2013,Hodille_PhysRevMat2018}.
Such complex emergent mechanisms require very large molecular dynamics simulations which in turn requires simple empirical potentials.

We develop a new potential for tungsten-hydrogen based on the embedded atom model, specifically designed for use with cascade simulations and high dose irradiation modelling. 
Our principle point of comparison for the correct behaviour of our potential is electronic structure Density Functional Theory (DFT) calculations.
In this work we report some new DFT calculations performed to find the relaxation volumes of vacancy-hydrogen clusters.
The accurate prediction of relaxation volumes for various induced defects using DFT calculations plays an important role in understanding lattice swelling and trapping mechanisms in materials under irradiation within the multi-scale modelling approach \cite{Nguyen_NIMB2015,Hofmann_Acta2015,Mason_JPCM2017,DeBacker_NucFus2018,Mason_2019,Wrobel_COMMAT2021}, and this work specifically aims to reproduce these properties with an empirical potential.
We also make extensive comparison to a empirical potential produced by Wang et al~\cite{Wang_JPCM2017}, which we believe gives the best values in the current literature for hydrogen-vacancy cluster binding energies.

In section \ref{sec:WW}, we show that the monovacancy and vacancy cluster properties in pure tungsten match DFT results, including the low divacancy binding energy~\cite{Muzyk_PRB2011}, vacancy migration barriers, and relaxation volumes.
In section \ref{sec:WH}, we show the binding energies of H atoms to vacancies and interstitials, and relaxation volumes of vacancy-H clusters also match DFT results.
We also find an increased binding of the H atom to the surface compared to previous work.
We show that molecular hydrogen is stable inside voids, and that increasing the H/point defect ratio increases the stability of open vacancy structures.
Finally in section \ref{sec:extended}, we compute properties for lattice defects typical of irradiation damage- small dislocation loops and voids.
We show that molecular H$_2$ exists within nanocavities, and that a supersaturation of hydrogen gas atoms can significantly disrupt the local atomic structure around dislocation loops in order to reduce energy.

Throughout this work we have made comparisons excluding the zero point energy contributions to the binding energy, which are not insignificant for hydrogen isotopes.
This choice has been made to avoid fitting the classical potential energy surface to the curvature of the potential energy found in DFT. 
This is also an acknowledgement of the fact that molecular dynamics (MD) simulations are classical, and so do not include quantum-mechanical phonon effects.
We have, however, computed zero point energy corrections for this potential and added these to the results separately as appropriate.
We have also decided to compute and report relaxed zero temperature configurations rather than dynamic effects at elevated temperatures, leaving this to future work.

This empirical potential is well suited to simulations of the retention of deuterium in highly irradiated tungsten~\cite{Mason_PRM2021}, and to investigate trap mutation mechanisms whose effect is clear experimentally~\cite{Simmonds_JNM2017,Simmonds_NF2022} but where the atomistic origin is currently speculation.

\subsection{The form of the potential}

The basic form of the embedded atom potential suggested by Daw and Baskes~\cite{Daw_PRB1984} gives the potential energy $E_i$ of atom $i$ as a local function of the positions and types of the neighbouring atoms. We write the distance from atom $i$ to a neighbour $j$ as $x_{ij}$, and the types of atoms $i$ and $j$ as $\alpha$ and $\beta$ respectively. Then
    \begin{equation}
        E_i = \frac{1}{2} \sum_{j} V_{\alpha\beta}\left( x_{ij} \right) + F_{\alpha}\left( \sum_{j} \phi_{\beta} \left( x_{ij} \right) \right),
    \end{equation}
where $V_{\alpha\beta}(x)$ is a pairwise interaction energy depending on the type of both atoms, $\phi_{\alpha} \left( x \right)$ is an electron density function, depending only on the type of the neighbour, and $F_{\alpha}(\rho)$ is an embedding function capturing the many-body nature of metallic bonding. 
This potential form for alloys is supported by the classical molecular dynamics (MD) code \texttt{LAMMPS}~\cite{LAMMPS} as the \texttt{eam/alloy} pair style.

We have made changes to the forms of the functions $V$, $\phi$, and $F$, suggested by Finnis and Sinclair~\cite{Finnis_PMA1984} and later developed by others for similar potentials.
The first observation we make is that potential hardening at short range is often required to simulate displacement cascades~\cite{Ackland_PMA1987, Bjorkas_NIMB2009, Becquart_JNM2021, Juslin_JNM2013}. Typically the short-range hardening used is the universal ZBL form~\cite{SRIM} added as a very large correction at short range. 
Here we start with the ZBL form for the pairwise interaction $V_{\alpha}$, and add a small correction near equilibrium lengths. 
Secondly, we note that in the second moment approximation~\cite{Ackland_JPhysF1988}, the embedding function $F_{\alpha}[\rho]$ has a square root form, which is very successful for modelling transition metals.
We add a spline correction to this function.
The final observation we make is that potentials need to be smooth in their second derivatives at least in order to avoid discontinuities in the quasiharmonic phonon spectrum. We ensure this by using quintic splines with well-separated knots.
The explicit form of the potential, and all the parameters needed to reconstruct a data-table is given in section \ref{sec:parameterization}.
  
\section{Fitting the potential}

The three parts of the potential, giving the W-W, H-H and W-H interactions, were fitted separately, then together, to reproduce sets of electronic-structure derived data.
Possible fits to the target data set were then run through an expanded set of more time-consuming tests, reported here, to check for transferability.
We report the overall best performing parameterization.

\subsection{W-W properties}
\label{sec:WW}
\begin{figure*}
    \centering
    \includegraphics[width=0.9\linewidth]{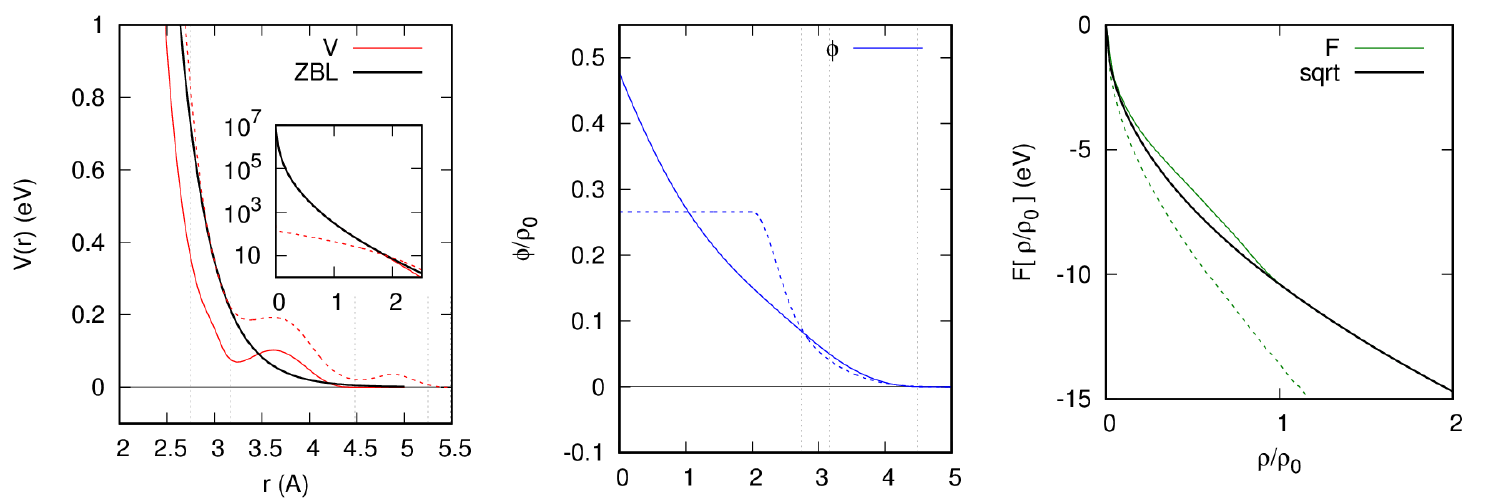}
    \caption{W-W potential comparison to ref~\cite{Marinica_JPCM2013} (dashed lines). a) pairwise potential, black line ZBL repulsion. b) density function, c) embedding function, black line square root form. The vertical lines in a) and b) indicates the equilibrium spacing of neighbours in the bcc lattice. The density is scaled in b) and c) by the level for a W atom in the perfect bcc lattice.
    }
    \label{fig:dimer_WW}
\end{figure*}

The W-W potential was fitted to structural and elastic properties of bcc tungsten, to vacancy cluster properties and surface formation energies, and to single interstitial properties. 
The initial fit was chosen to be as close as possible to the MNB potential~\cite{Mason_JPCM2017}- a modification of the transferable Finnis-Sinclair form with improved vacancy cluster properties- within the constraints of the new functional form.
The downhill simplex method was used to minimise the fitting in a least-squares sense.
Additional weighting was placed on keeping the knot points in the potential far apart, in order to mitigate overfitting.
The fitted properties are given in table \ref{tab:WW}, and include formation energies (denoted $E_f$ in the text and tables), relaxation volumes ($\Omega_{\rm{rel}}$), binding energies ($E_b$), and migration energies ($E_m$) for point defects.
A comparison is given to two empirical EAM forms: `EAM-2' given by Marinica et al~\cite{Marinica_JPCM2013}, as this is the basis for the W-W part of both the Bonny \cite{Bonny_JPCM2014} and Wang~\cite{Wang_JPCM2017} W-H potentials; and our earlier MNB potential~\cite{Mason_JPCM2017}.
We find that the potential described here is good in these targetted properties, as indeed are the comparison potentials generally. A direct comparison between the functional forms for this work and the EAM-2 potential~\cite{Marinica_JPCM2013} is shown in figure \ref{fig:dimer_WW}. 
The only significant difference to the eye is that this work has a shorter range- the EAM-2 potential has a long range repulsion between equilibrium atom spacings which has a strong effect on the energetics of interstitial defects.

\begin{figure}
    \centering
    \includegraphics[width=0.9\linewidth]{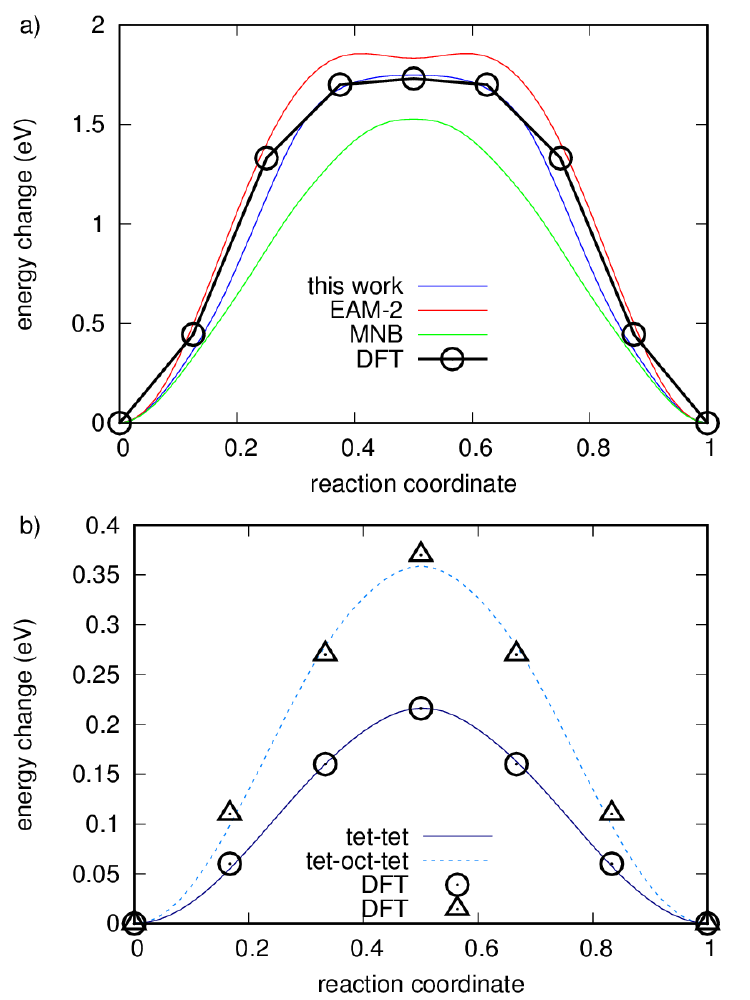}
    \caption{NEB calculation for migration in W. a) monovacancy migration, compared to DFT result from \cite{Heinola_NF2018}, with lines to guide the eye. b) H in otherwise pure W. DFT comparison results from Kong et al \cite{Kong_Acta2015}.
    }
    \label{fig:migration}
\end{figure}

We first consider vacancy and open surface properties. 
The new potential performs comparably to the MNB in the vacancy cluster and surface energy properties, which was itself designed to improve on the EAM-2 results. The only significant improvement we make in this redraft of the W-W part of the potential is a better monovacancy migration energy, which was included in the refitting.
We find with a nudged elastic band calculation that this is raised from 1.52 eV in MNB to 1.75 eV in this work, which can be compared to the DFT value 1.75 eV~\cite{Mason_JPCM2017}.
This result is illustrated in figure \ref{fig:migration}a).

The divacancy energies and migration barriers are also a good fit to the DFT values of ref~\cite{Mason_JPCM2017}. 
If one vacancy is at lattice site \hkl[000], the lowest energy configuration is with the second vacancy at nearest neighbour \hkl[\half \half \half], though even here the binding energy is only 0.080 eV.
To move the second vacancy to position \hkl[100], \hkl[110], \hkl[111] has a barrier 1.900, 2.146, 1.696 eV respectively, which can be compared to the pattern of the DFT values 1.825, 1.807, 1.717 eV~\cite{Mason_JPCM2017}. 
The trivacancy binding energy is reproduced well, but its migration barrier is not. 
This potential finds the migration barrier 1.752 eV, similar to that of the monovacancy. 
But DFT is able to rehybridise the electrons at the saddle point configuration, and finds a low saddle point of 1.146 eV.
This problem with EAM empirical potentials was discussed in ref~\cite{Mason_JPCM2017}.

\begin{table*}[]
    \centering
    \begin{tabular}{cccccccc}
        Category    & property          &   units       &   DFT             &   expt            &   this work   &   EAM-2       &   MNB         \\
        \hline
        perfect     &   lattice const   &   \AA         &   3.186$^{a}$     &   3.1652$^{b}$    &   3.145       &   3.140       &   3.1652 \\
        bcc         &   cohesive energy &   eV          &                   &   8.90$^{b}$      &   8.949       &   8.900       &   8.900 \\
        lattice     &   $C_{11}$        &   eV/\AA$^3$  &   3.229$^{a}$     &   3.324$^{c}$      &   3.201       &   3.260       &   3.222 \\       
                    &   $C_{12}$        &   eV/\AA$^3$  &   1.224$^{a}$     &   1.279$^{c}$      &   1.257       &   1.265       &   1.263 \\         
                    &   $C_{44}$        &   eV/\AA$^3$  &   0.888$^{a}$     &   1.018$^{c}$      &   1.020       &   0.996       &   0.998  \\
        point       &   $E_f^{v}$       &   eV          &   3.619$^{d}$     &                   &   3.631       &   3.49        &   3.727 \\
        defects     &   $\Omega_{\rm{rel}}^{v}$  &   $\Omega_0$  &   -0.32$^{a}$     &                   &   -0.329      & \emph{-0.045} &   -0.36 \\
                    &   $E_m^{v}$       &   eV          &   1.756$^{d}$     &                   &   1.754   &   1.856        &   1.523 \\
                    &   $V_2$ binding   &   eV          &   0.048$^{d}$     &                   &   0.080       & \emph{0.490}  &   0.170 \\
                    &   $V_3$ binding   &   eV          &   0.311$^{d}$     &                   &   0.365       & \emph{1.417}  &   0.433 \\
                    &   $E_f^{i}([111])$ &   eV         &   10.5            &                   &   9.736       &   10.52       &   9.31     \\
                    &   $\Omega_{\rm{rel}}^{i}$   &   $\Omega_0$  &   1.57$^{a}$      &                   &   1.485       & \emph{1.172}  &   1.40 \\
                    &   $E_f^{i}([100]) - E_f^{i}([111])$  &   eV  & 0.29   &                   &   0.236       &   0.30        &   0.25  \\
        surface     &   $E_{\{110\}}$   &   eV/\AA$^2$  &   0.212       &                   &   0.220       & \emph{0.144}  &   0.218 \\
        energy      &   $E_{\{100\}}-E_{\{110\}}$ & eV/\AA$^2$ & 0.047  &                   &   \emph{0.028}       &   \emph{0.026}       &   \emph{0.021}
    \end{tabular}
    \caption{Targetted properties of the W-W potential. A comparison is given to the empirical EAM form `EAM-2' \cite{Marinica_JPCM2013}, and a our earlier MNB potential \cite{Mason_JPCM2017}.    
    $^{a}$ ref \cite{Mason_JAP2019},
    $^{b}$ ref \cite{Finnis_PMA1984},
    $^{c}$ ref \cite{Featherston_PR1963},
    $^{d}$ ref \cite{Mason_JPCM2017}
    In this table and subsequently, italics indicate values which could produce significant errors in simulation. These values are discussed in the text.
    }
    \label{tab:WW}
\end{table*}

We define the formation and binding energies of a $m$- vacancy cluster in tungsten as, respectively,
    \begin{eqnarray}
        E_f^{v_m} &=&  E(v_m) - (N-m)E_0       \nonumber\\
        E_b^{v_m} &=& m E_f^{v_1} - E_f^{v_m},
    \end{eqnarray}
where $E(v_m)$ is the lowest total energy of a box containing $N$ lattice sites of which $m$ are vacant, and $E_0$ is the energy per tungsten atom ( equal to the negative of the cohesive energy ).
For this work, the lowest energy has been found in the same manner as ref~\cite{Mason_JPCM2017}. 
$m$ sites were selected as a string of nearest neighbour positions, within a box of $7 \times 7 \times 7$ conventional unit cells, removing the atoms on those sites, and relaxing atomic positions with constant volume. Up to one thousand randomly selected strings of atoms were compared to find a good minimum energy configuration. The relaxed energy quoted includes an elastic energy correction~\cite{Varvenne_PRB2013}. 

\begin{figure}
    \centering
    \includegraphics[width=0.9\linewidth]{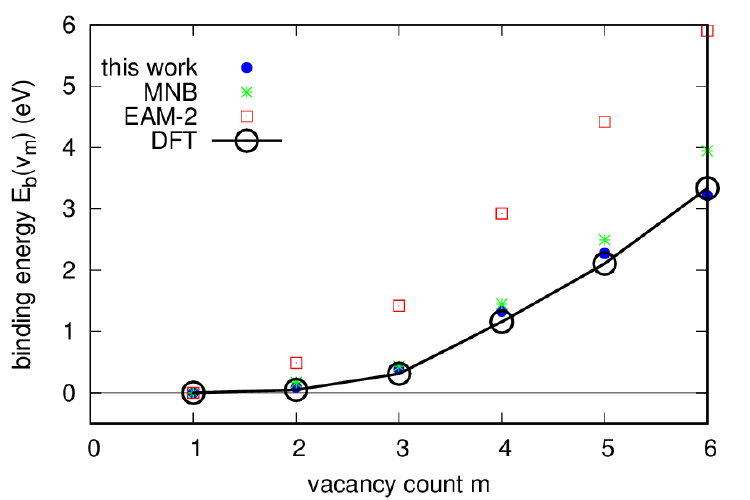}
    \caption{(Total) binding energy of vacancy clusters in pure tungsten.}
    \label{fig:VmH0}
\end{figure}

\begin{figure}
    \centering
    \includegraphics[width=0.9\linewidth]{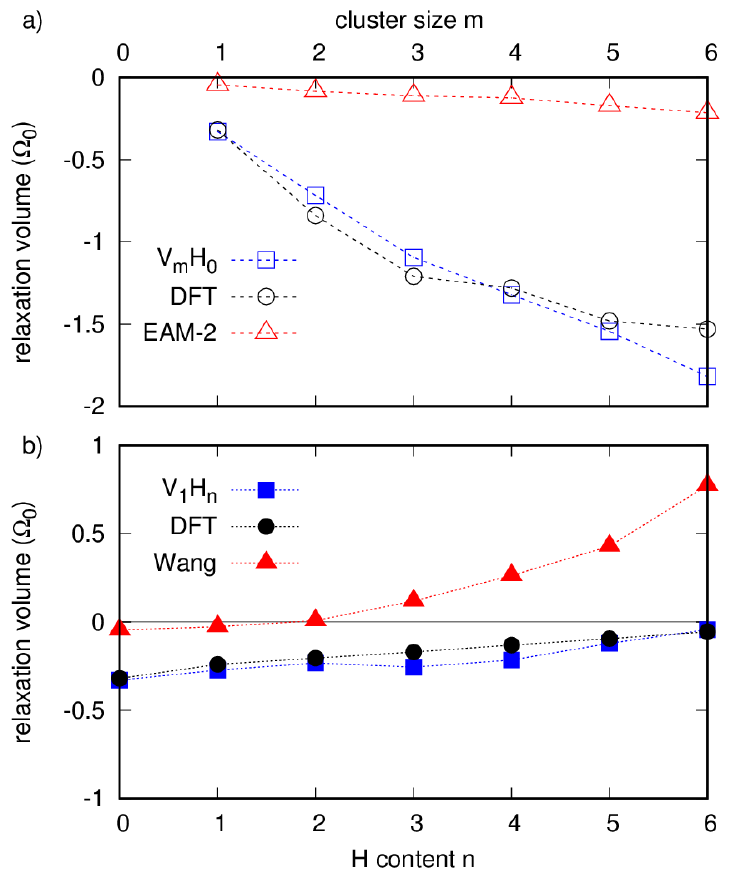}
    \caption{Relaxation volumes of a) empty vacancy clusters $V_m$ and b) monovacancies filled with H $V_1 H_n$, in units of the atomic volume $\Omega_0$. The lines are to guide the eye. DFT results in figure a) from \cite{Mason_JPCM2017}.
    }
    \label{fig:relaxation_volumes}
\end{figure}

The binding energy of vacancy clusters in pure W is shown in figure \ref{fig:VmH0}, and relaxation volumes of vacancy clusters are shown in figure \ref{fig:relaxation_volumes} a).
We see that the current potential finds excellent agreement with DFT values from~\cite{Mason_JPCM2017}, a slight improvement on the previous MNB potential and a considerable improvement on the EAM-2 potential.

The properties of some unreconstructed low index surfaces in pure W are given in table \ref{tab:Wsurface}. The surface energies and stresses are a good match to DFT results from ref~\cite{Ma_PRM2020}. 
As noted in ref~\cite{Mason_JPCM2017}, the high formation energy of the surfaces in bcc tungsten is due to the pseudo-gap in the electronic density of states near the Fermi level - the practical outcome of which is to make electron-deficient tungsten atoms higher energy than the square-root form of the embedding function second-moment ( Finnis-Sinclair ) potentials would suggest. The approximate energy level of the surfaces is therefore adjusted by reducing the embedding energy in this low-electron-density region, as can be seen in figure \ref{fig:dimer_WW}c).
We note that the energy of the \hkl[100] surface is lower than the \hkl[111] surface, in common with other empirical EAM or tight-binding second-moment approximation ( Finnis-Sinclair ) potentials. 
In these potentials, the leading energy contribution can be estimated by simply counting bonds: the surfaces \hkl(110), \hkl(211), \hkl(100) and \hkl(111) have, respectively a nearest- and next-nearest- neighbour count of (6,4), (5,3), (4,5), and (4,3). Hence the \hkl(110) surface has the lowest energy, in qualitative agreement with DFT calculations, but the \hkl(111) surface has the highest energy.
However the empirical potential models that relate the surface energies to the cohesive energies are flawed because while they include the contributions from the d band in transition metals, they ignore the general trend of sp electrons to spread smoothly at the surface. 
The latter sp-d hybridisation not only affects the attractive bonding contributions but also the repulsive energy from electrostatic and exchange-correlation contributions. Using DFT calculations, the predicted surface energies can be improved by 10-20$\%$ leading to the different trends observed \cite{Methfessel1992}.

The practical result of this failing of EAM potentials is that the \hkl(100) surface is the least reliable, and both the surface energy and surface stress are not well reproduced.

\begin{table}[]
    \centering
    \begin{tabular}{cccccc}
        property        &   DFT             &   expt       &   this work     &  EAM-2\\
        \hline
        $\gamma_{110}$  &   0.212           & 0.229 &   0.220   &   \emph{0.144}   \\
        $\gamma_{211}$  &   0.224           &   &   0.248   &   \emph{0.171}   \\    
        $\gamma_{111}$  &   0.239           &   &   0.266   &   \emph{0.184}   \\
        $\gamma_{100}$  &   0.259           &   &   0.248   &   \emph{0.170}   \\
        $s_{110}$       &   0.301,0.178     &   &   0.363,0.194 \\
        $s_{211}$       &   0.223,0.151     &   &   0.220,0.151 \\
        $s_{111}$       &   0.135,0.135     &   &   0.130,0.130 \\
        $s_{100}$       &   0.147,0.147     &   &   \emph{0.242},\emph{0.242}
    \end{tabular}
    \label{tab:Wsurface}
    \caption{Surface properties of the W potential. Surface energies $\gamma_{hkl}$ and surface stresses $s_{hkl}$ for surface normal \hkl[hkl] are given in eV/$\AA^2$. DFT values from ref~\cite{Ma_PRM2020}, experiment from ref~\cite{Tyson_SurfSci1977}.
    Values showing significant discrepancy are indicated with italics.
    }
\end{table}


We now consider interstitial properties of the W-W potential.
Interstitial properties were computed in a $5 \times 5 \times 5$ conventional unit cell with fixed volume and the elastic correction made, to be a close comparison to DFT values.
The formation energies for single interstitials in \half\hkl[111] and \hkl[100] orientations are raised slightly compared to the previous MNB potential, which we have achieved by adding a small repulsive bump in the pairwise potential $V(r)$ between second- and third- neighbour positions, similar to the one seen in EAM-2.
The relaxation volume of the \half\hkl[111] crowdion is found to be 1.485 times the volume per tungsten atom in the perfect 0 K lattice, $\Omega_0$. 
This is a good match to the DFT- determined relaxation volume, 1.57 $\Omega_0$~\cite{Mason_JAP2019}.

\begin{figure}
    \centering
    \includegraphics[width=0.9\linewidth]{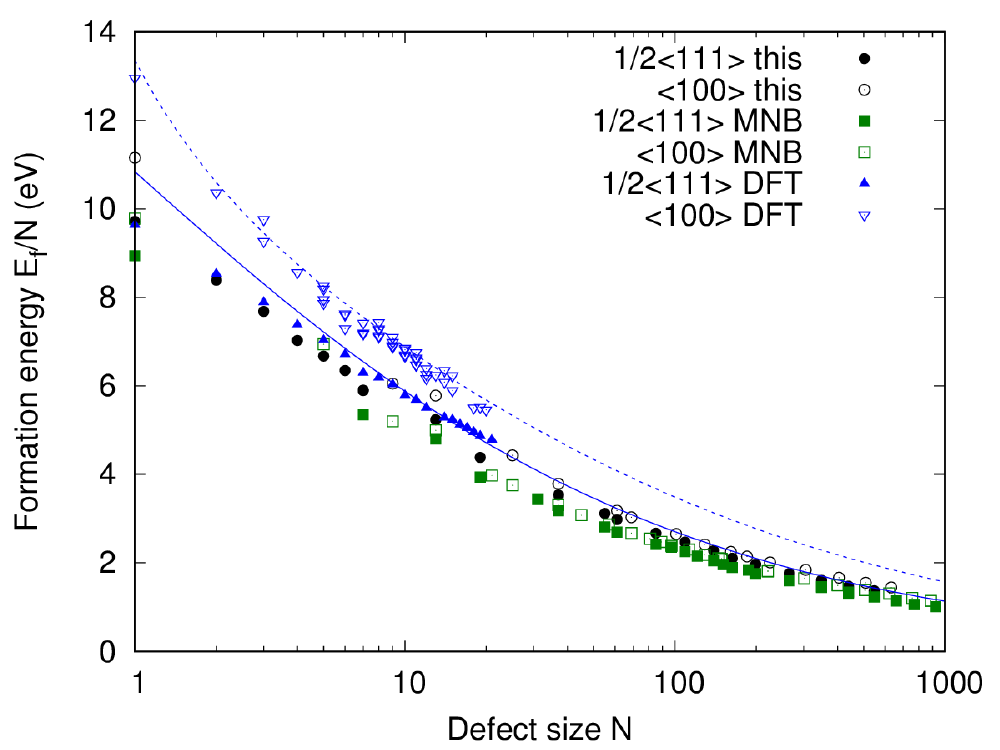}
    \caption{Formation energy of interstitial clusters and loops per point defect. DFT results from ref~\cite{Alexander_PRB2016}, with extrapolating lines being fits to the data from ref~\cite{Mason_JAP2019}.}
    \label{fig:loops}
\end{figure}

The transferability of the potential to larger interstitial clusters is tested by computing the formation energy of interstitial clusters and loops in pure W, shown in figure \ref{fig:loops}. 
Small interstitial clusters up to size 6 point defects were generated using a method similar to vacancy clusters, constructing a string of $m$ lattice sites at random, then placing an interstitial atom into randomly selected tetrahedral or octahedral sites along the string. 500 such strings were constructed to find a low energy, using a cell size $16 \times 16 \times 16$ unit cells with fixed volume.
Larger interstitial loops were constructed to be circular within a box size $48 \times 48 \times 48$ unit cells with fixed volume.
The energy per point defect is slightly higher than in the MNB or AT potentials, and the increase in energy of \hkl[100]- Burgers vector defects over those with \half\hkl[111] is raised slightly. A comparison to DFT~\cite{Alexander_PRB2016, Mason_JAP2019} shows both these changes are improvements, and in particular the \half\hkl[111] loops are in good agreement with DFT. We note that the GAP potential of ref~\cite{Byggmastar_PRB2019} is closer to DFT for the \hkl[100] loops. 
Some other tungsten empirical potentials (~\cite{Derlet_PRB2007,Marinica_JPCM2013}) find \hkl[100] loops have the lower energy. 
While the significance of this ordering is not clear in atomistic high dose microstructures generated by MD~\cite{Gra20,Gra21,Derlet_PRM2020}, we shall see in section \ref{sec:WH} that the energy ordering of dislocation loops is further affected by the presence of hydrogen isotopes, so these potentials cannot be used as a base for seeking trap mutation mechanisms depending on defect energy ordering.
        
\begin{figure}
    \centering
    \includegraphics[width=0.9\linewidth]{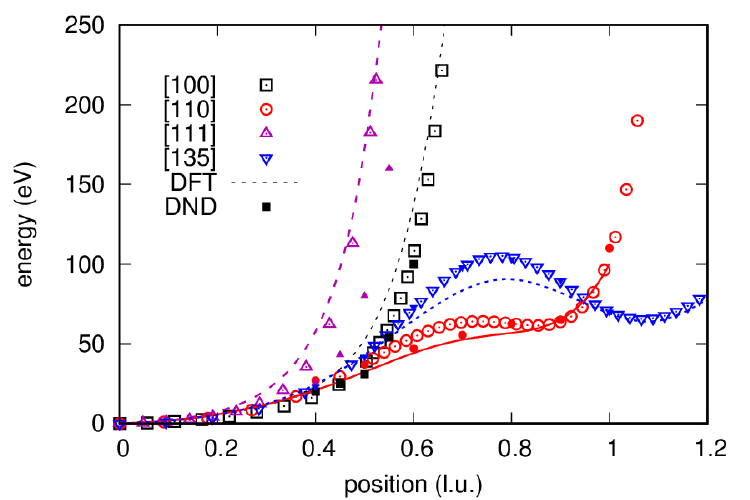}
    \caption{Energy for quasi-static drag of a single atom in direction indicated. Lines denote DFT values from ref~\cite{Becquart_JNM2021}. Some representative points for the DND interatomic potential from ref~\cite{Becquart_JNM2021} are shown using filled symbols.}
    \label{fig:qsd}
\end{figure}

We consider the transferability of the short-range part of the W-W potential using the Quasi-Static Drag (QSD) method introduced by Becquart et al~\cite{Becquart_JNM2021}.
A single atom is displaced along a fixed vector direction, and the energy is recorded without relaxing the atoms.
We consider vector directions \hkl[100], \hkl[110], \hkl[111], \hkl[135].
As the atoms get close together, the potential tends to the ZBL repulsive form, and the energy rises rapidly.
The results are shown in figure \ref{fig:qsd}.
DFT results generated in ref~\cite{Becquart_JNM2021} using semi-core electrons in the PAW approximation are indicated in the figure, and we see this work follows the DFT values quantitatively for the head-on collisions considered (\hkl[100] and \hkl[111]), and is acceptable for the \hkl[110] and \hkl[135] directions.
To make a proper comparison with literature, we include indicative results generated using the DND W potential~\cite{Derlet_PRB2007}, known to generate collision cascades in agreement with experimental observations~\cite{Yi_EPL2015}.
We see this work is very similar to DND, closer than the other empirical potentials considered in ref~\cite{Becquart_JNM2021}, but we note that this work follows DFT more closely even than DND for the head-on \hkl[111] collision.

%

\subsection{H-H properties}    

\begin{figure*}
    \centering
    \includegraphics[width=0.9\linewidth]{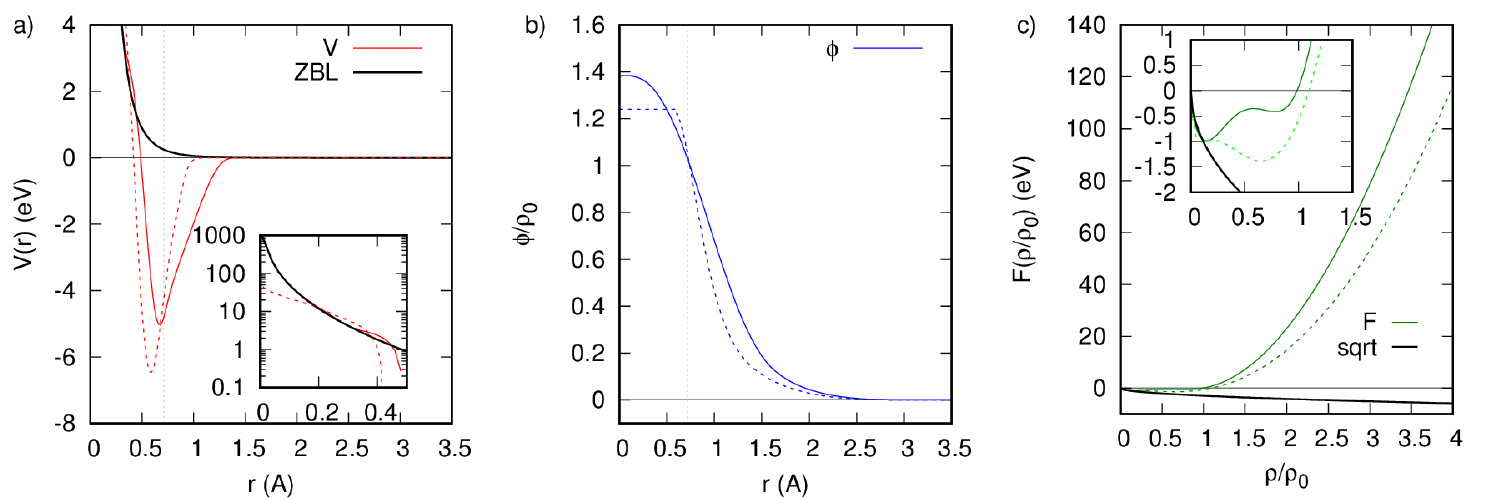}
    \caption{H-H potential comparison to ref~\cite{Wang_JPCM2017} (dashed lines). a) pairwise potential, black line ZBL repulsion. b) density function, c) embedding function. The vertical line in a) and b) indicates the dimer equilibrium spacing. The density is scaled in b) and c) by the level for the dimer at equilibrium spacing.
    }
    \label{fig:dimer_HH}
\end{figure*}

Though much is known about molecular hydrogen, classical empirical potentials can only reproduce a rather limited subset of their properties. 
The initial fit was chosen to be as close as possible to Wang~\cite{Wang_JPCM2017}, and a comparison of properties is given in table \ref{tab:HH}.
There is little difference between the two EAM potentials, in particular note that the H$_2$ dimer is formed using either potential, and H$_3$ and close packed (hcp) hydrogen are strongly unfavoured. A direct comparison between the functional forms for this work and the Wang potential\cite{Wang_JPCM2017} is shown in figure \ref{fig:dimer_HH}. 
The only small change made here is increasing the binding energy at low electron density- this change allows slightly higher binding energy of the H atom to a surface. 
Note that while the embedding function technically starts with a square root form, the deviation from it is substantial.
The majority of the work done fitting of the H-H potential was tuning the embedding function $F[\rho]$ in order to reproduce W-H binding energies.

\begin{table}[]
    \centering
    \begin{tabular}{ccccc}
        property        &   units           &   expt       &   this work    &   Wang       \\
        \hline
        $H_2$ length   &   \AA              &   0.74       &   0.744        &   0.726      \\
        1/2 $E_f^{H_2}$  &   eV             &   -2.26      &   -2.121       &   -2.358      \\
        $H_2$ spring const   & eV/\AA$^2$   &   35.95      &   34.44        &   31.0       \\
        1/3 $E_f^{H_3}$  &   eV             &              &   -0.808       &   -1.389       \\
        cohesive E hcp   &   eV             &              &   0.990        &   1.386      \\
    \end{tabular}
    \caption{Properties of the H-H potential. A comparison is given to the empirical Wang potential \cite{Wang_JPCM2017}.    
    }
    \label{tab:HH}
\end{table}

\subsection{W-H properties}    
\label{sec:WH}

\begin{figure}
    \centering
    \includegraphics[width=0.9\linewidth]{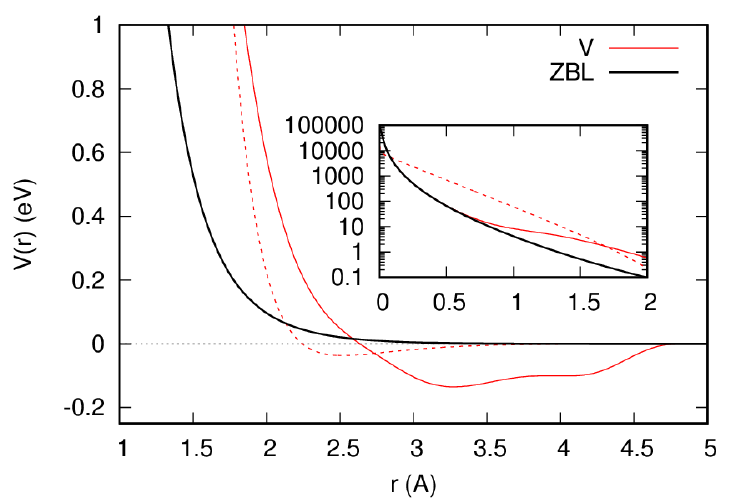}
    \caption{W-H potential comparison to ref~\cite{Wang_JPCM2017} (dashed lines). Black line shows ZBL repulsion. 
    }
    \label{fig:dimer_WH}
\end{figure}

The W-H interaction used an initial fit as close as possible to Wang et al.
A direct comparison of the potential forms for this work and Wang is shown in figure \ref{fig:dimer_WH}.
This work shows stronger binding at typical H-W separation distances, and weaker repulsion at short range.

The targetted properties are given in table \ref{tab:WH}.
Some of the values have zero point energy contributions included. For this work, these have been computed by finding the phonon frequencies $\{ \omega_i \}$ directly from the full dynamical matrix, and computing the sum $E_{ZPE} = \sum_i \half \hbar \omega_i$.
The stable position for interstitial H is the tetrahedral site, with the octahedral site 0.36 eV higher in energy, in agreement with DFT.
The migration barriers for a mobile interstitial H atom moving from one site to the next are illustrated in figure \ref{fig:migration}b), and also are a good match to DFT.
We have also targetted unrelaxed interstitial formation energies, a good test of the W-H potential at short bond length.

\begin{table*}[]
    \centering
    \begin{tabular}{ccccccc}
        Category    & property          &   units       &   DFT             &   expt            &   this work   &   Wang            \\
        \hline
        H in bcc W  &   $E_f^{H,tet}$   &   eV          &    0.93$^{a}(1.04)$     &  1.04 $\pm$ 0.17  &   0.798(1.049)    &   1.056         \\
                    &   $E_f^{H,oct}-E_f^{H,tet}$ & eV  &    0.44$^{a}$     &                   & 0.359  &   0.35            \\
                    &   $E_m^{H,tet-tet}$ & eV  &    0.21$^{a}$     &                   & 0.216  &    0.22         \\
                    &   relax vol H tet &   $\Omega_0$  &   0.15$^{b}$           &                   &   0.133       &   \emph{0.241}    \\
                    &   $E_b^{H-H}$(2nn)&   eV          &   -0.446$^{c}$     &                   &  -0.468 &   -0.525      \\
                    &   $E_b^{H-H}$(3nn)&   eV          &   -0.093$^{c}$     &                   &  -0.013       &   -0.112      \\
                    &   $E_f^{H,tet}(unrelaxed)$   &   eV          &    2.87     &     &   2.682      &   \emph{0.968}        \\
                    &   $E_f^{H,oct}(unrelaxed)$   &   eV          &    2.81     &     &   2.629      &   3.217        \\
        H in vac    &   $E_b^{v-H}$     &   eV          &   1.28(1.43)$^{e}$  &                &   1.233(1.290)       &   1.011            \\
                    &   $E_b^{vH-H}$    &   eV          &   1.25(1.41)$^{e}$  &                &   1.349(1.456)        &   1.098            \\
                    &   $E_b^{vH_2-H}$  &   eV          &   1.11(1.22)$^{e}$  &                &   0.975(1.032)        &   0.819            \\
                    &   $E_b^{vH_3-H}$  &   eV          &   1.00(1.11)$^{e}$  &                &   1.050(1.074)        &   0.688            \\
                    &   $E_b^{vH_4-H}$  &   eV          &   0.91(1.00)$^{e}$  &                &   0.705(0.742)        &   0.561            \\
                    &   $E_b^{vH_5-H}$  &   eV          &   0.32(0.47)$^{e}$  &                &   0.486(0.670)        &   0.285            \\
                    & $E_b^{H \rm sub}$ &   eV          &                      &                &   0.375        &   0.550            \\
                    & $\Omega_{\rm{rel}}^{vH}$   & $\Omega_0$ & -0.24       &   &  -0.273        &  \emph{-0.027}   \\
                    & $\Omega_{\rm{rel}}^{vH_2}$ & $\Omega_0$ & -0.205      &   &  -0.232        &  \emph{0.008}    \\
                    & $\Omega_{\rm{rel}}^{vH_3}$ & $\Omega_0$ & -0.170      &   &  -0.256        &  \emph{0.119}    \\
                    & $\Omega_{\rm{rel}}^{vH_4}$ & $\Omega_0$ & -0.131      &   &  -0.216        &  \emph{0.263}    \\
                    & $\Omega_{\rm{rel}}^{vH_5}$ & $\Omega_0$ & -0.094      &   &  -0.121        &  \emph{0.429}    \\
                    & $\Omega_{\rm{rel}}^{vH_6}$ & $\Omega_0$ & -0.055      &   &  -0.043        &  \emph{0.774}    \\                    
        H + int     & $E_b^{i-H}$       &   eV          &   0.33$^b$        &    &   0.357         &  0.416         \\
        H + surface & $E_b[100]$        &  eV           & 0.93$^f$    & 0.7,0.82$^f$  &  \emph{0.312}\emph{(0.466)}   &   \emph{0.146}     \\
    \end{tabular}
    \caption{Targetted properties of the W-H potential. Values in brackets include zero point energy. A comparison is given to the empirical EAM potential by Wang et al \cite{Wang_JPCM2017}. 
    $^{a}$ ref~\cite{Fernandez_ActaMater2015}  
    $^{b}$ ref~\cite{DeBacker_NucFus2018}
    $^{c}$ ref~\cite{Liu_JNM2009}
    $^{d}$ ref~\cite{Qin_JNM2015}
    $^{e}$ ref~\cite{Heinola_PRB2010b}       
    $^{f}$ ref~\cite{Johnson_JMR2010}.
    The binding energy for a pair of H interstitial atoms is given for tetrahedral sites separated by \hkl[^1\!/\!_2\, 0\, 0] and \hkl[^1\!/\!_2\, ^1\!/\!_4\, ^1\!/\!_4].
    }
    \label{tab:WH}
\end{table*}

 The present DFT calculations of relaxation volumes were performed for this work using the VASP \textit{ab initio} simulation code, using the PAW method~\cite{PhysRevB.47.558,PhysRevB.54.11169,KRESSE199615} with semi-core electrons included through the use of pseudo-potentials.
 It is important to emphasize that the inclusion of semi-core electrons in the valence states has a significant effect on the predicted formation energies of both vacancy and self-interstitial atom (SIA) defects for all the bcc transition metals~\cite{NguyenManh_PRB2006,Nguyen_JMS2012,Hofmann_Acta2015}. 
In the present study of vacancy interaction with hydrogen atoms, the exchange-correlation effects were described using the Perdew-Burke-Ernzerhof generalised gradient approximation~\cite{Perdew_PRL1996}. A kinetic energy cut-off of 500 eV was used, with a $5\times 5\times 5$ Monkhorst-Pack grid for electron density k-points employed in the case with super-cell ( $5 \times 5 \times  5$ ) calculations. 
The set of hydrogen-vacancy defect clusters used was similar to those described in~\cite{Heinola_PRB2010,Ohsawa_JNM2007} with larger super-cell size. 
The full cell relaxation method was used to evaluate the relaxation volumes. A comparison between fully relaxed calculations with those using the constant volume approximation in a combination with the corrected elastic dipole tensor calculations has been recently discussed in~\cite{Wrobel_COMMAT2021}. While the latter approximation can be extended to investigate the relaxation volume of radiation induced defects at mesoscopic scale in pure tungsten~\cite{Mason_2019}, it has been demonstrated clearly in the case of interaction between helium atoms and vacancy clusters that the former method is more reliable not only in reproducing experimental data of the lattice swelling but also the modulus change in a helium-implanted tungsten alloys.

Relaxation volumes of vacancy clusters and monovacancies filled with H are shown in figures \ref{fig:relaxation_volumes}a) and \ref{fig:relaxation_volumes}b) respectively.
The new potential fitted here is a good match to DFT, suggesting it should well reproduce the local stresses generated by high concentrations of dissolved hydrogen gases and vacancy clusters.
The relaxation volume of an H atom bound to a crowdion is found to be 1.593 $\Omega_0$, in good agreement with the DFT value 1.719 $\Omega_0$.

\begin{figure}
    \centering
    \includegraphics[width=0.9\linewidth]{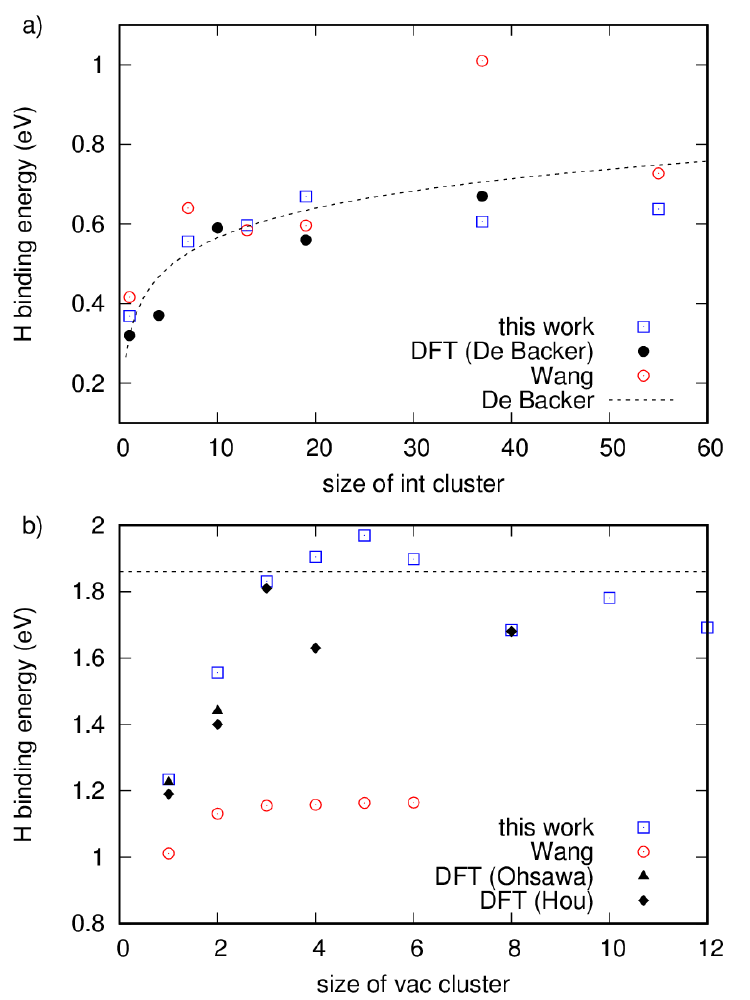}
    \caption{Binding energy of single hydrogen atoms to extended defects in tungsten. a) Binding energy to interstitial clusters and \half\hkl<111> loops. The dashed line indicates the fit suggested in ref~\cite{DeBacker_NucFus2018}. b) Binding energy to vacancy clusters. The dashed line indicates the binding energy of the \hkl[001] surface from ref~\cite{Heinola_PRB2010}. DFT values from De Backer et al~ref\cite{DeBacker_NucFus2018}, Hou et al~\cite{Hou_NatMat2019}, and Ohsawa et al~\cite{Ohsawa_JNM2007} }
    \label{fig:singleH}
\end{figure}

We test transferability by considering the binding of single H atoms to defect clusters, and the binding of multiple H atoms to point defects, as both these have comparable DFT calculations in the literature.
Figure \ref{fig:singleH}a) shows the highest binding energy of a single H atom to interstitial clusters and \half\hkl<111> interstitial loops.
The binding energy increases from 0.36 eV for the single SIA to 0.64 eV for a large loop ( m=55 interstitials, diameter 2 nm ), in good accord with the DFT calculations of De Backer et al~\cite{DeBacker_NucFus2018}.
Figure \ref{fig:singleH}b) shows the highest binding energy of a single H atom to vacancy clusters.
The binding energy increases from 1.23 eV for the monovacancy to plateau around 1.7 eV for large vacancy clusters. 
This increase in binding energy was reported in a DFT study by Hou et al~\cite{Hou_NatMat2019}, and is consistent with the high binding energy to a surface in the limit of infinite void size.
The potential of Wang does not reproduce a significant increase in binding energy of H with the size of the vacancy cluster.
Care should be taken overestimating the importance of this result in dynamic studies, as the binding energy of H to any void-space is high, so instead of finding a true thermodynamic equilibrium, MD simulations will most likely be stuck in a long-lived transient state with H atoms decorating the first vacancy cluster they encounter and rarely being detrapped.

\begin{figure}
    \centering
    \includegraphics[width=0.9\linewidth]{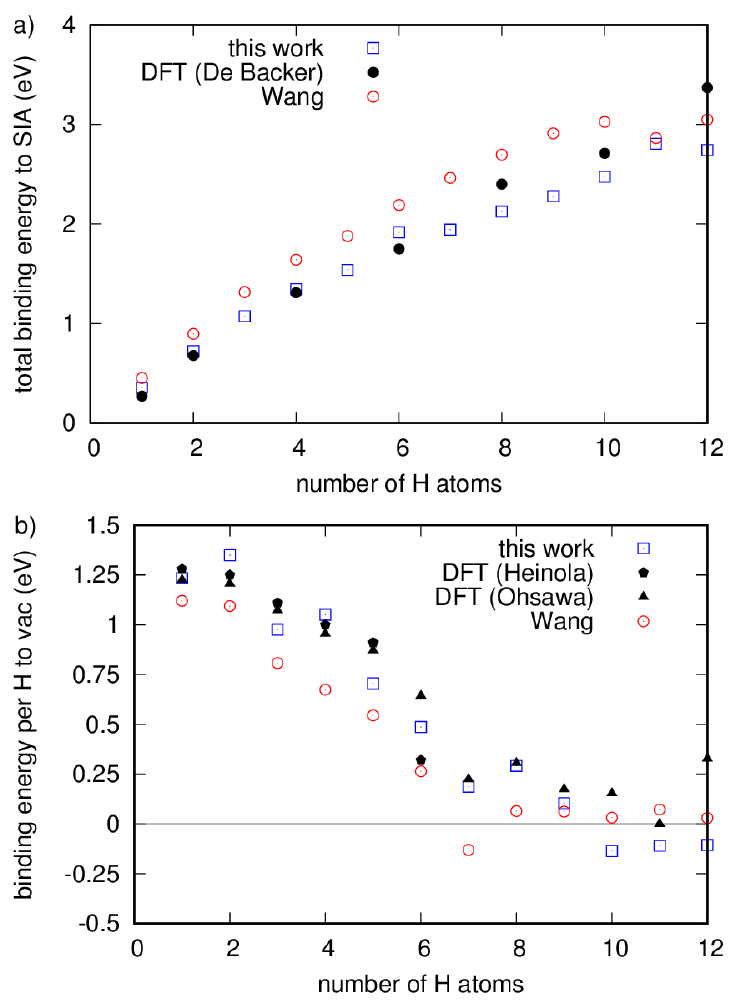}
    \caption{Binding energy of multiple hydrogen atoms to point defects in tungsten. a) Total binding energy to single SIA (crowdion). b) Total binding energy to monovacancy. DFT values from De Backer et alref~\cite{DeBacker_NucFus2018}, Heinola et al~\cite{Heinola_PRB2010}, and Ohsawa et al~\cite{Ohsawa_JNM2007} }
    \label{fig:pointDefectW}
\end{figure}

Figure \ref{fig:pointDefectW}a) shows the binding energy of multiple H atoms to a single interstitial, with the DFT comparison from ref~\cite{DeBacker_NucFus2018}. Both the Wang potential and this work do a similarly good job reproducing the near-linear increase in total binding energy up to 12 H atoms.
Figure \ref{fig:pointDefectW}b) shows the binding energy per H atom to a single vacancy.
Again, both empirical potentials are showing the correct trend, with this work producing a slightly higher binding energy, closer to DFT.
The H atoms are placed at the octahedral \hkl[\half 0 0] interstices.

Binding energies of an H atom to the surface are given in table \ref{tab:WHsurface}. 
Note that the reference point for these energies is the vacuum level of molecular hydrogen.
These energies are somewhat below that predicted by DFT and found experimentally, with the highest binding energy being 0.487 eV for an H atom on the \hkl(110) surface. 
Care should be therefore taken when attempting to model the equilibrium between gas phase and the surface, particularly if the \hkl[100] surface is used. 
Note that the Wang potential has poor surface energy properties generally, with surface energies 30\% lower than DFT, and that potential only gives a positive binding energy for H on a \hkl[100] surface.
The improvement made here is modest, but is the most we have been able to achieve -- the near-field atomic configuration of an H atom on the surface looks similar to the atomic configuration inside a vacancy.
In order to stabilise H on a surface further with an EAM potential without changing the vacancy binding energy requires a low-electron-density binding state ( which can be seen in figure \ref{fig:dimer_HH}c) ), but be were not able to make this state even more binding without destabilising molecular H$_2$.

\begin{table}[]
    \centering
    \begin{tabular}{cccccc}
        property        &   DFT             &   expt       &   this work    &   Wang \\
        \hline
        \hkl[110]  &  0.75        &   &   0.487(0.722)  &  -0.291    \\
        \hkl[211]  &          &   &   0.389(0.707)   &  -0.024     \\    
        \hkl[111]  &          &   &   0.394(0.696)   &   0.001     \\
        \hkl[100]  &  0.93    & 0.7,0.82  &  0.312(0.466)   &   0.146     \\
    \end{tabular}
    \caption{Surface binding energies of H on W, computed as $E_f( \rm{surface} ) + 1/2 E_f( H_2 ) - E_f( \rm{surface}:H)$.
    DFT and experimental values reported in ref~\cite{Johnson_JMR2010} .
    Figures in brackets include zero point energy contributions. 
    }
    \label{tab:WHsurface}
\end{table}

\section{Large defect clusters in a hydrogen-rich atmosphere}
\label{sec:extended}

In this section we consider how the energy levels of prototypical nanoscale lattice defects produced by radiation damage~\cite{Jenkins2001} are affected by the binding of multiple hydrogen isotope atoms.
There is little DFT data for larger clusters, so the results in this section are predictions made with the current empirical potential, together with a discussion of where they may prove important for gas retention studies.

%

\begin{figure}
    \centering
    \includegraphics[width=0.6\linewidth]{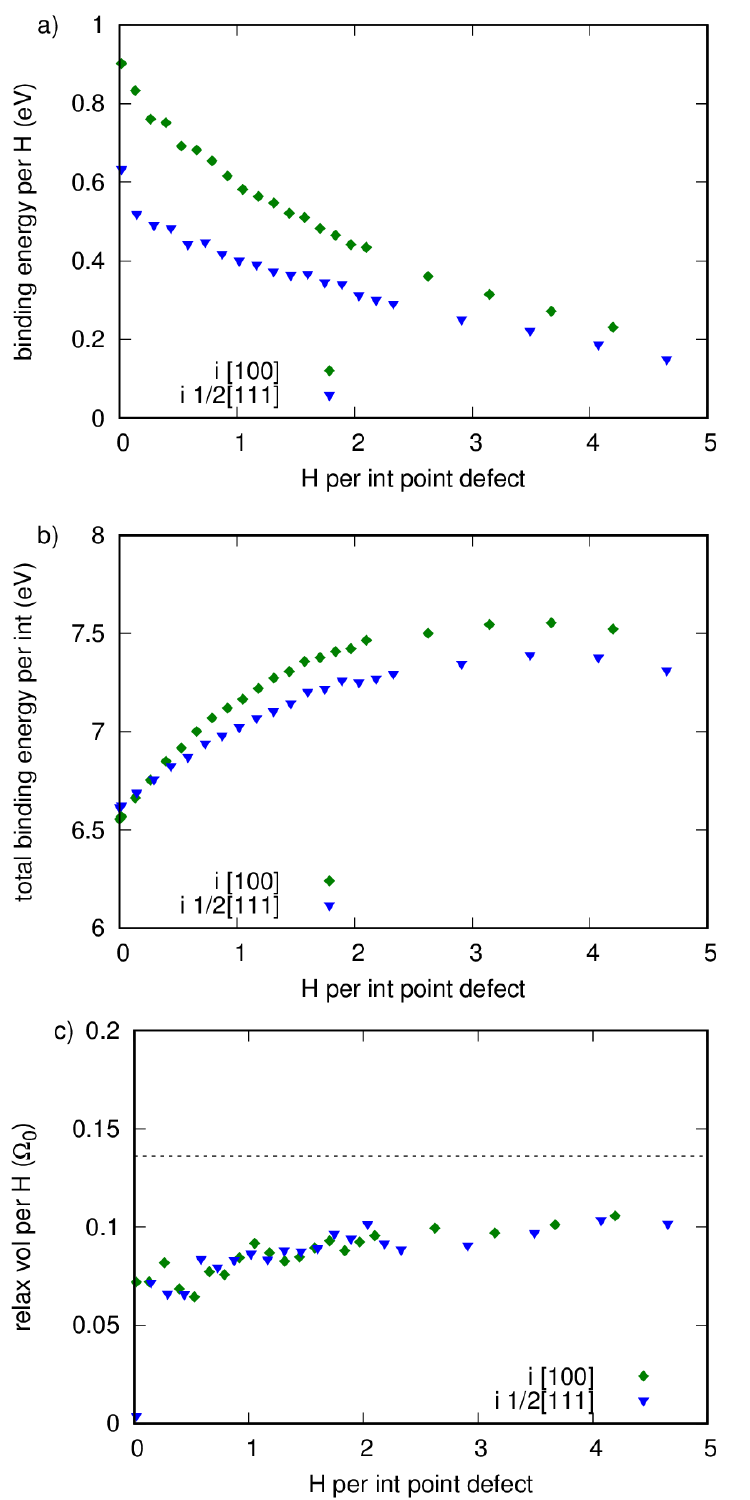}
    \caption{Binding energy of hydrogen atoms to extended interstitial-type defects in tungsten as a function of the ratio of H atoms to point defects. a) Binding energy per H atom to interstitial clusters and dislocation loops. b) Total binding energy per interstitial ( binding energy of interstitial defect and binding energy of H atoms ) for extended interstitial defects. c) Relaxation volume per bound H atom, expressed as a fraction of the volume per tungsten atom $\Omega_0$. Dashed line shows relaxation volume of a single H atom in the tetrahedral position in perfect W.}
    \label{fig:extendedDefects_int}
\end{figure}

In figure \ref{fig:extendedDefects_int}a) we plot the binding energy of multiple H atoms to interstitial dislocation loops, in terms of binding energy of H atoms to an existing loop, ie the quantity $E_b( I_m H_n ) = E_f( I_m H_0 ) + n E_f(H) - E_f( I_m H_n )$, where $E_f(I_m H_n)$ is the formation energy of the $m$ interstitial loop decorated with $n$ H atoms.
In this figure we have created circular dislocation loops with radius $3 a_0$, giving a diameter of about 2 nm -- around 60 interstitials.
In both the \hkl<100> and \half\hkl<111> Burgers vector cases, the H atoms are bound close to the dislocation core. At higher ratios of H atoms to interstitials, excess H atoms are weakly bound in the region of tensile stress just outside the disc of inserted atoms. The binding energy is greater to the \hkl[100] loop than to the \half\hkl[111] loop, as the former has the larger Burgers vector and so creates a greater tensile stress.
Figure \ref{fig:extendedDefects_int}b) replots the data as the total binding energy bringing all interstitials and H atoms from infinity, ie the quantity $E_b^{\mathrm{tot}}( I_m H_n ) = (N+m) E_0 + m E_f( I_1 H_0 ) + n E_f(H) - E_f( I_m H_n )$, where the simulation cell has $N$ bcc lattice sites.
Though the undecorated \hkl<100> loop is higher energy than the \half\hkl<111> loop, adding H atoms brings the total formation energy of the \hkl<100> loop down faster, giving a crossover point between them at $m/n \sim 0.25$. 
As noted above, in this empirical potential, the energy gap between \hkl<100> and \half\hkl<111> is smaller than that predicted by DFT, so we should expect this crossover point to be higher in reality.
The qualitative result that dissolved hydrogen isotope gas should stabilise \hkl<100> loops is likely to be robust.

In figure \ref{fig:extendedDefects_int}c) shows the relaxation volume per H atom, computed from the dipole tensor and the elastic constants~\cite{Dudarev_Acta2017}. We see that each H atom contributes a small positive volume change, smaller than that of the interstitial H in a tetrahedral position in the perfect lattice. This volume change is in addition to the much greater relaxation volume of the interstitial loop itself, order $m \Omega_0$, indicating that an H atom brought from infinity to the loop will reduce the total elastic stress, but only slightly.
There is little difference between loops with \hkl<100> and \half\hkl<111> Burgers vectors. This result suggests little is to be gained in terms of reducing elastic stress build up by binding H atoms to interstitial defects, and coupled with the low binding energy per gas atom seen in figure \ref{fig:extendedDefects_int}a), we conclude that using this potential at room temperature and above there will be little hydrogen gas associated with interstitial loops except under plasma loading conditions. This is in line with the conclusions of the meso-scale study in ref~\cite{DeBacker_NucFus2017}.

\begin{figure}
    \centering
    \includegraphics[width=0.6\linewidth]{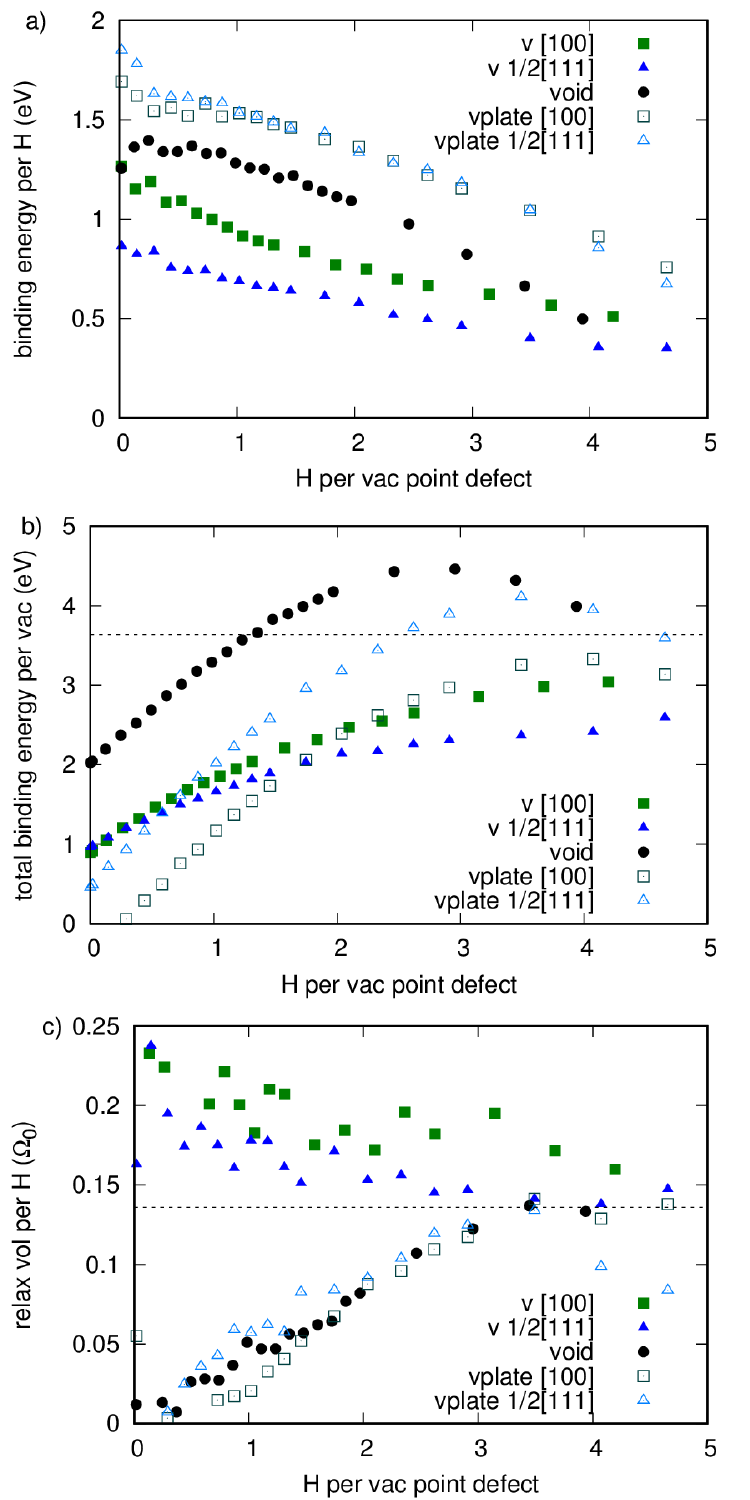}
    \caption{Binding energy of hydrogen atoms to extended vacancy-type defects in tungsten as a function of the ratio of H atoms to point defects. a) Binding energy per H atom to extended vacancy defects. b) Total binding energy per vacancy ( binding energy of vacancy defect and binding energy of H atoms ) for extended vacancy defects. Dashed line is the monovacancy formation energy. c) Relaxation volume per bound H atom, expressed as a fraction of the volume per tungsten atom $\Omega_0$. Dashed line shows relaxation volume of a single H atom in the tetrahedral position in perfect W.}
    \label{fig:extendedDefects_vac}
\end{figure}

Figure \ref{fig:extendedDefects_vac}a) shows the binding energy of H atoms to vacancy-type defects, $E_b( V_m H_n ) = E_f( V_m H_0 ) + n E_f(H) - E_f( V_m H_n )$.
Here we have considered the circular dislocation loops with Burgers vectors \hkl<100> and \half\hkl<111>, as well as the spherical void and the `open' vacancy loop~\cite{Gilbert_JPCM2008} - a platelet of vacancies with Burgers vector \hkl<100> and \half\hkl<111> formed when the surfaces inside the dislocation loop are created by separating the atomic planes. All defects have around 60 unoccupied lattice sites. As with the binding to interstitial loops, we see H atoms trapped near the vacancy loop dislocation cores, and with higher binding energies than the interstitial counterparts. This binding energy is slowly reduced as the ratio of H to vacancies is increased.
The open defects, the void and the vacancy platelets, have higher binding energy than the dislocation loops.
It is notable that the vacancy platelets have higher binding even than the void, which is likely due to a single H atom being able to stabilise the electron-deficient W atoms on either side of the platelet.

In figure \ref{fig:extendedDefects_vac}b) we replot the data using the total binding energy of the defect plus the binding of the H atoms per vacant lattice site, $E_b^{\mathrm{tot}}( V_m H_n ) = (N-m) E_0 + m E_f( V_1 H_0 ) + n E_f(H) - E_f( V_m H_n )$.
Here we see that the void is by far the strongest bound defect per vacancy at this defect size, which is to be expected as the energy of the open defects scales with surface area.
At high H occupation, $n/m>1$, it appears the binding energy per vacancy is greater than the formation energy per vacancy.
This does not mean that a void can spontaneously form- that would require the emission of interstitials which are very high energy point defects.
At low H occupation, $n/m \ll 1$, we see the \half\hkl<111> vacancy dislocation loop is the second most stable defect, with the \hkl<100> loop and platelet close behind at this defect size.
But as with the interstitial loop case, increasing the number of H atoms changes this order, and the \hkl<100> platelet is more stable with only a few added H atoms present. At higher ratios $n/m > 0.6$ , the \hkl<111> platelet becomes more stable than the dislocation loop, and at $n/m > 2$ , the \hkl<100> platelet becomes more stable than the loop.
These observations from the empirical potential, suggest that in a rich hydrogen isotope atmosphere condition, such as might be found in a plasma loading condition, there is a driving force for transformation between dislocation loops and open platelets.  

The relaxation volume per H atom bound to vacancy defects is shown in figure \ref{fig:extendedDefects_vac}c). 
In contrast to the result for interstitial loops, here we see considerably more structure.
H atoms bound to dislocation loops show a greater relaxation volume than in the tetrahedral interstitial position, which acts to alleviate the large negative relaxation volume ( order $- m \Omega_0$ ) of the vacancy loop itself. 
For open-volume defects - the void and plates considered here - the relaxation volume is very small for the first few gas atoms binding to the surface of the defect, as these do not contribute significant additional surface stress.
This result shows that non-zero stress applied as boundary conditions, or as locally varying stresses due to nearby defects, will further change the energy levels of the defects. 
In particular this result suggests that under tensile stress conditions, produced by interstitial defects in the early stages of radiation damage~\cite{Mason_PRL2020}, open structures may be further favoured. 
Further study of this effect is beyond the scope of this paper, but warrants future investigation.

\begin{figure*}
    \centering
    \includegraphics[width=0.22\linewidth]{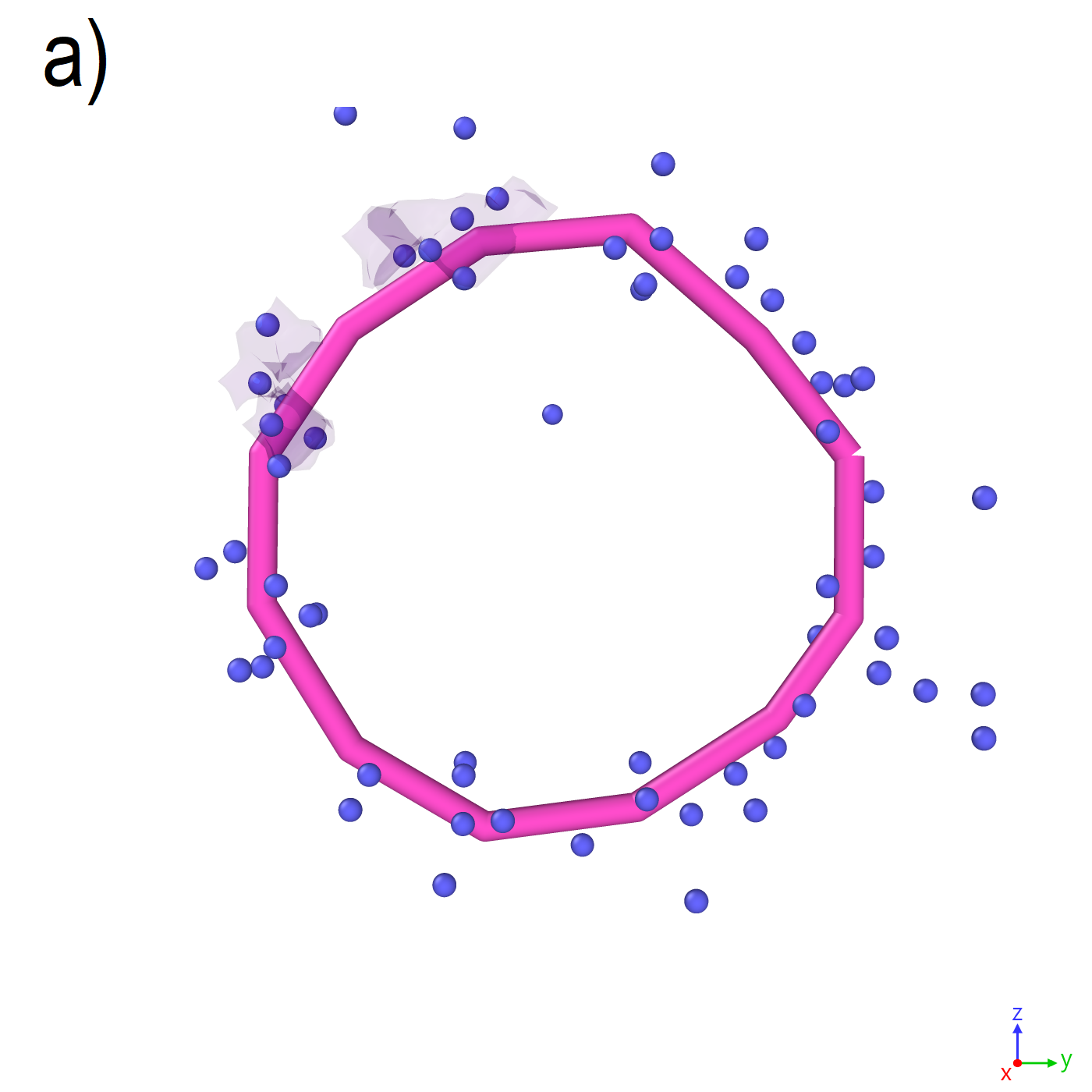}
    \includegraphics[width=0.22\linewidth]{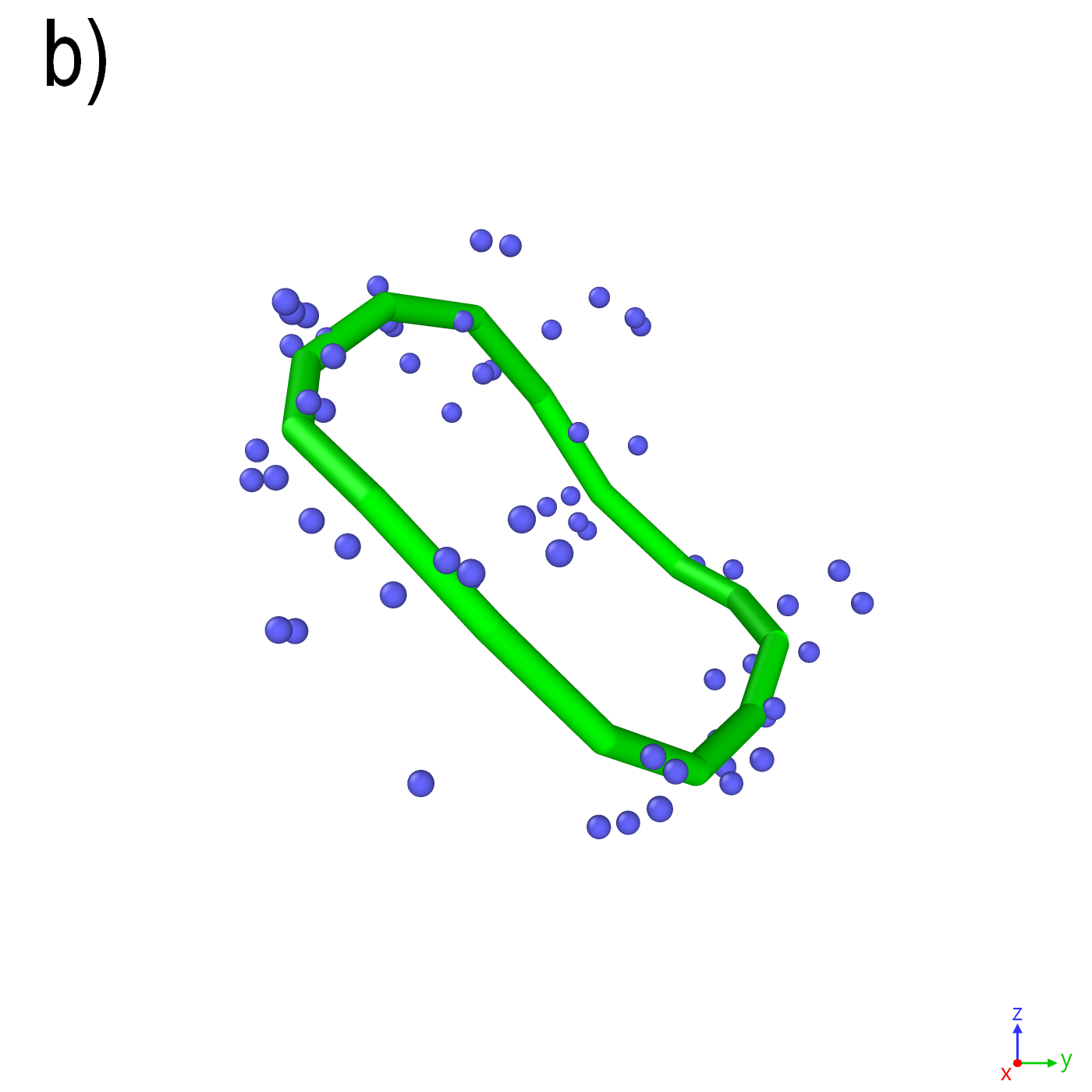}
    \includegraphics[width=0.22\linewidth]{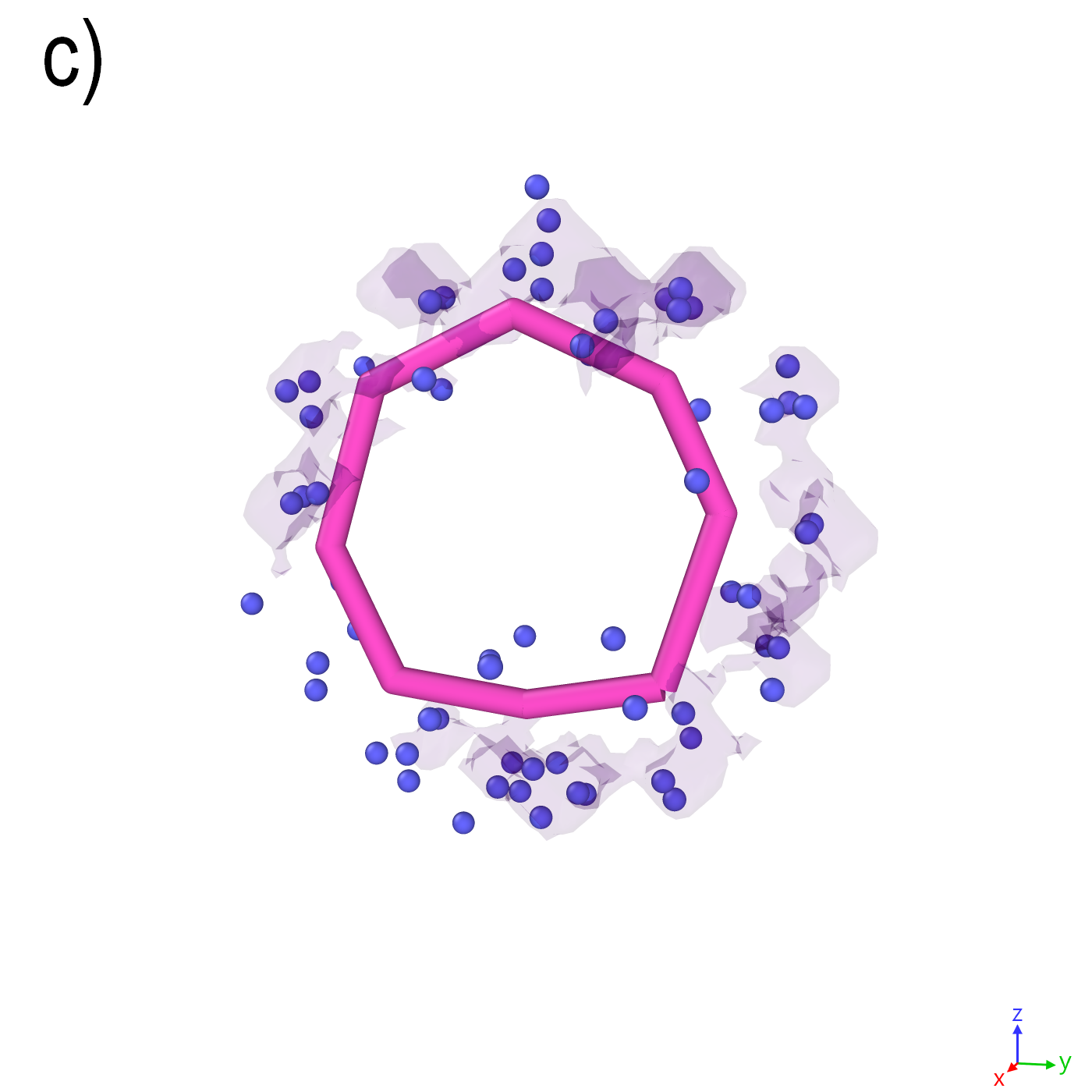}
    \includegraphics[width=0.22\linewidth]{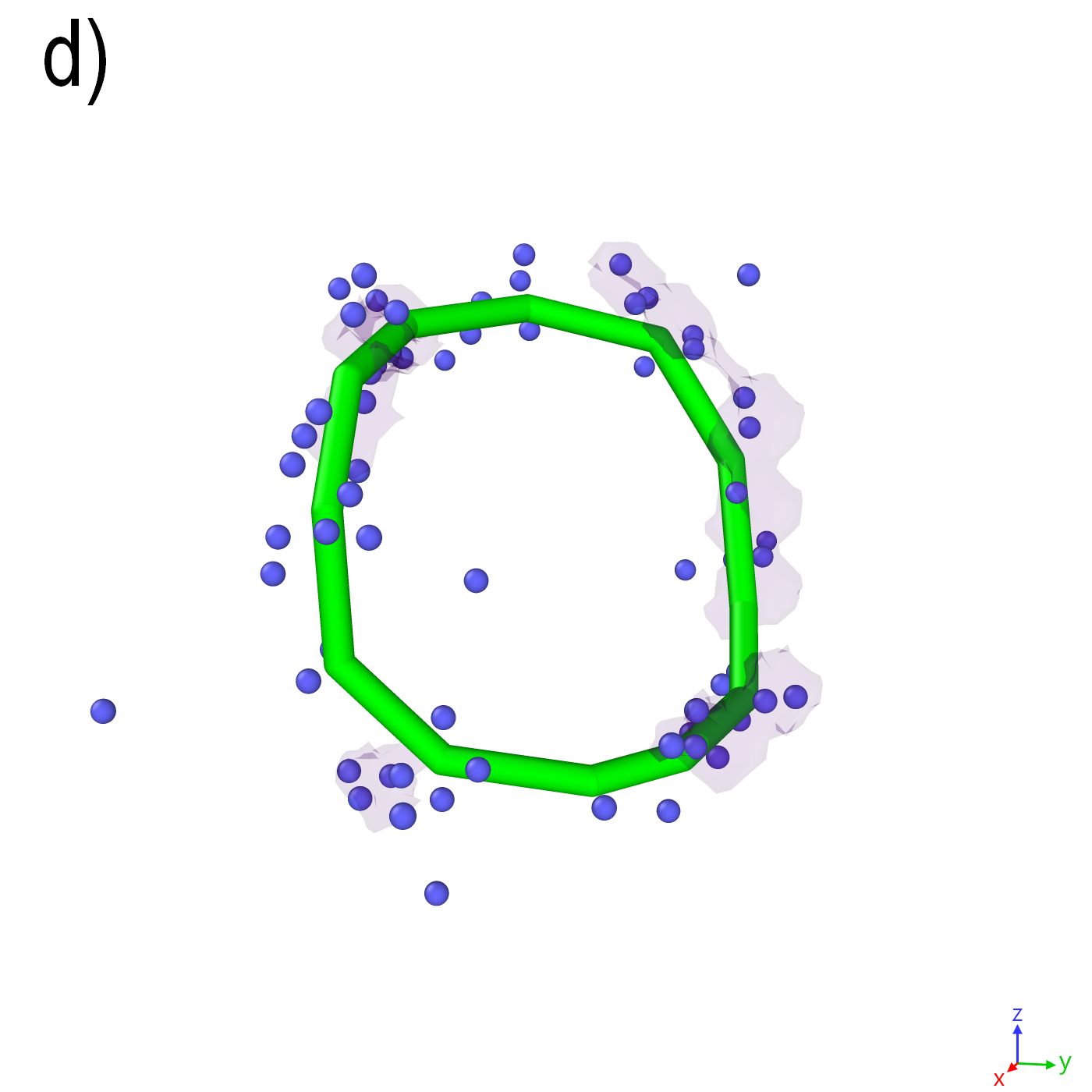}    \\
    \includegraphics[width=0.22\linewidth]{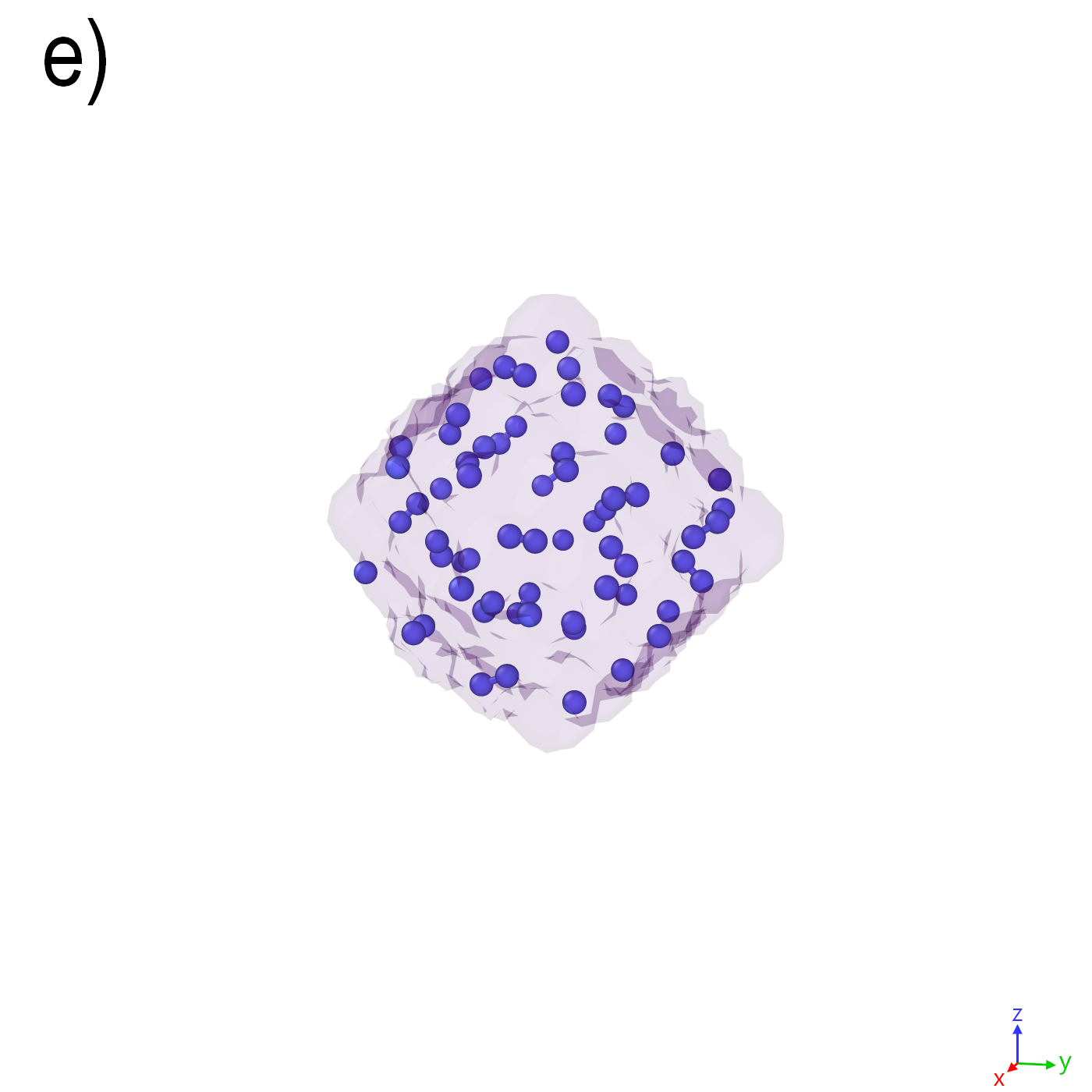}
    \includegraphics[width=0.22\linewidth]{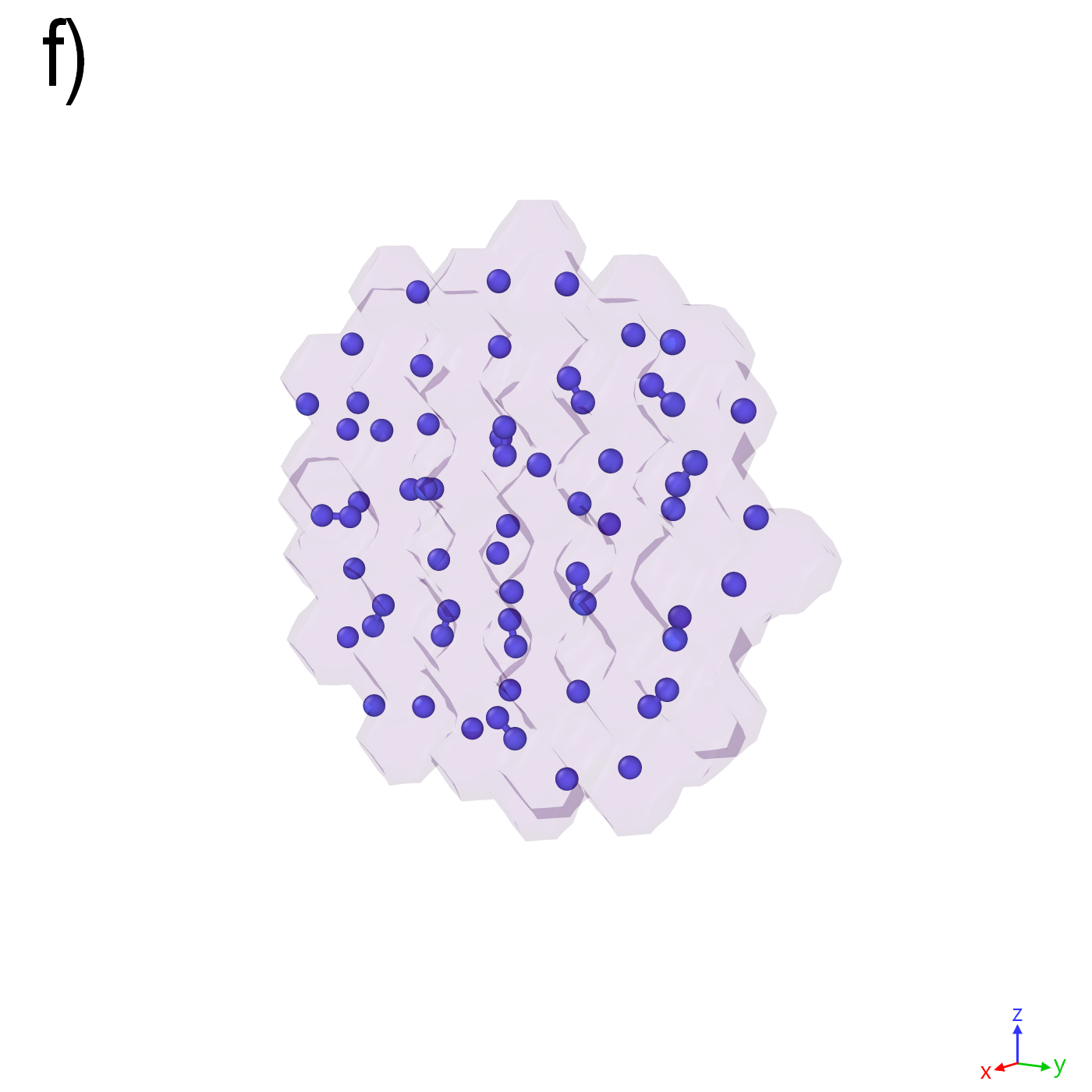}
    \includegraphics[width=0.22\linewidth]{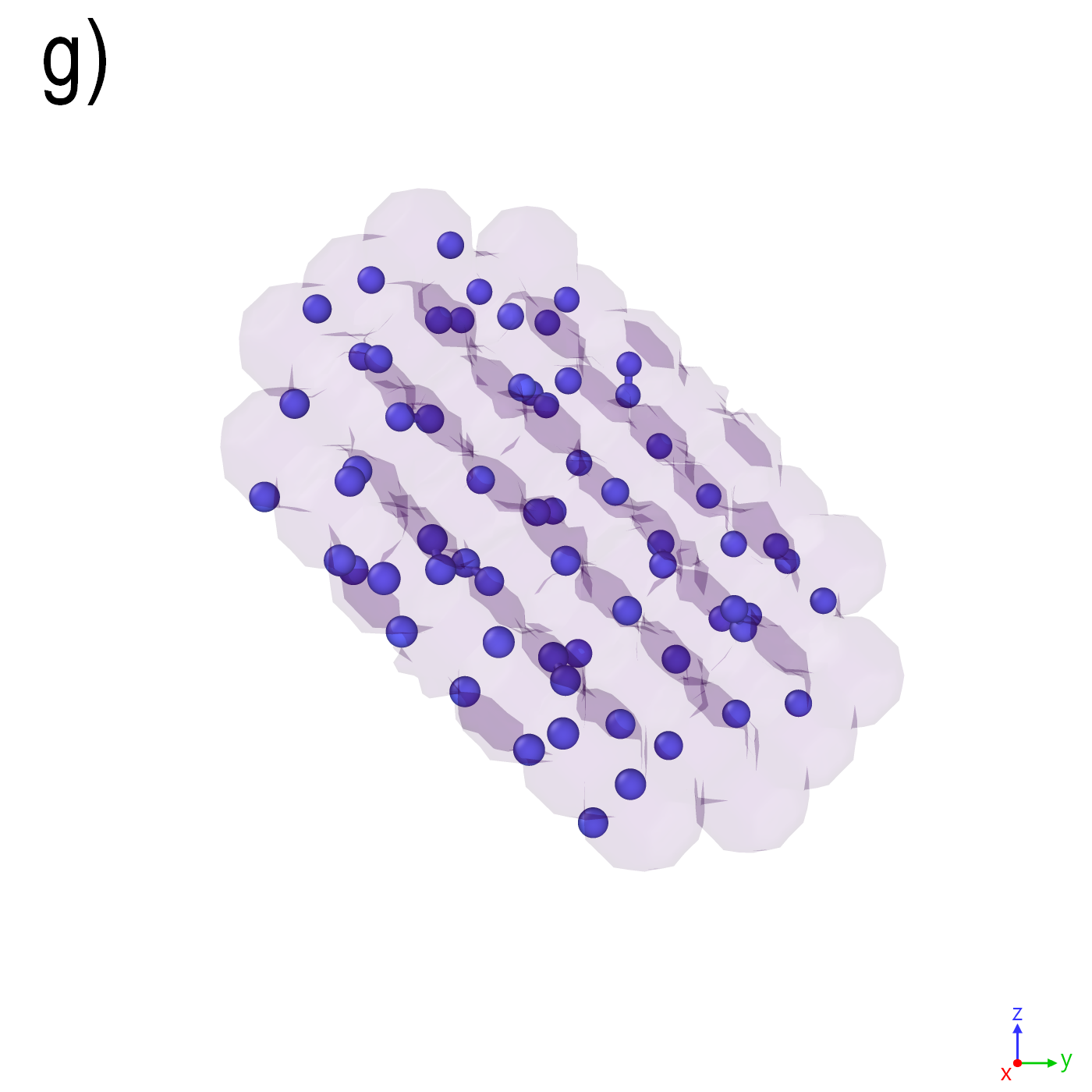}        \\
    \includegraphics[width=0.22\linewidth]{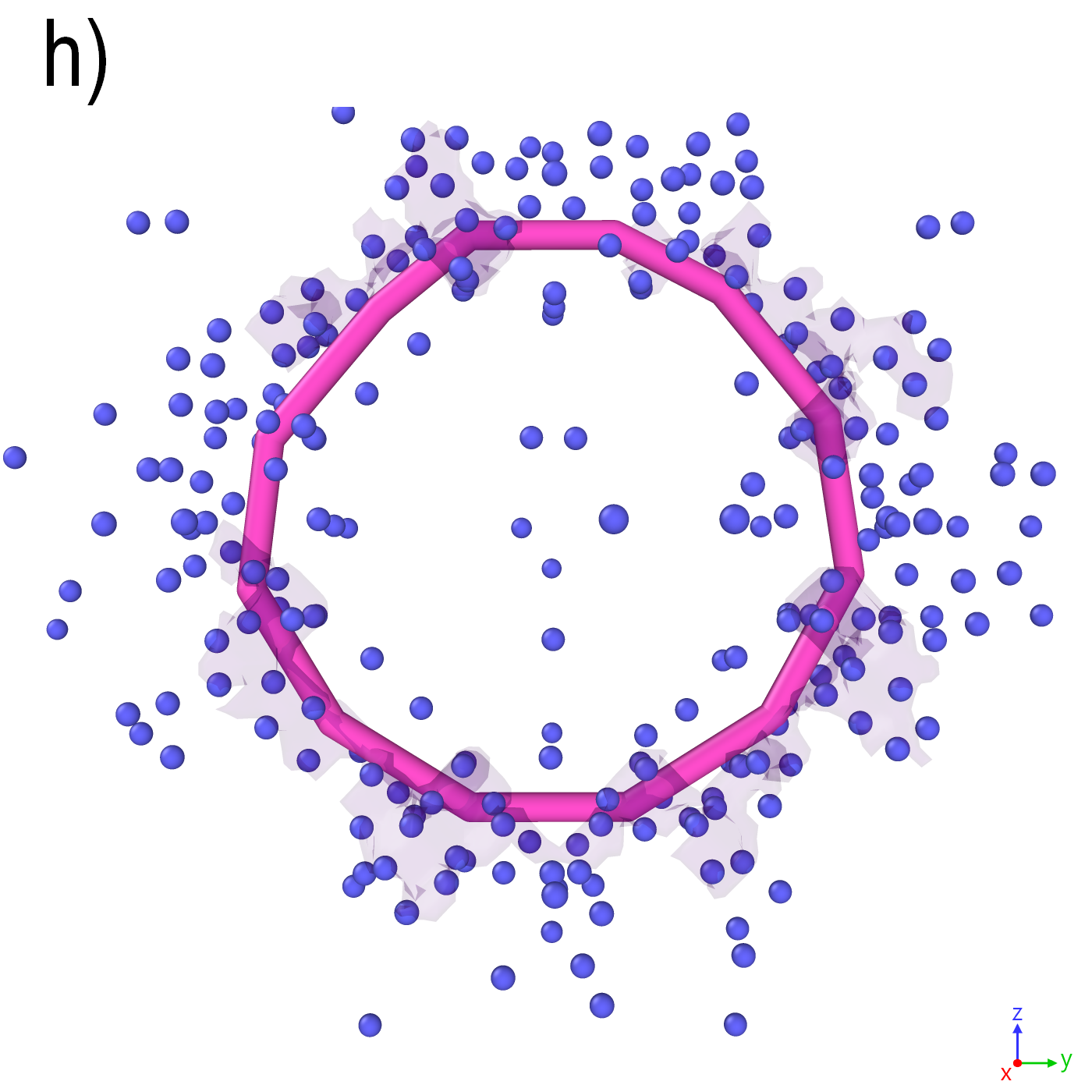}
    \includegraphics[width=0.22\linewidth]{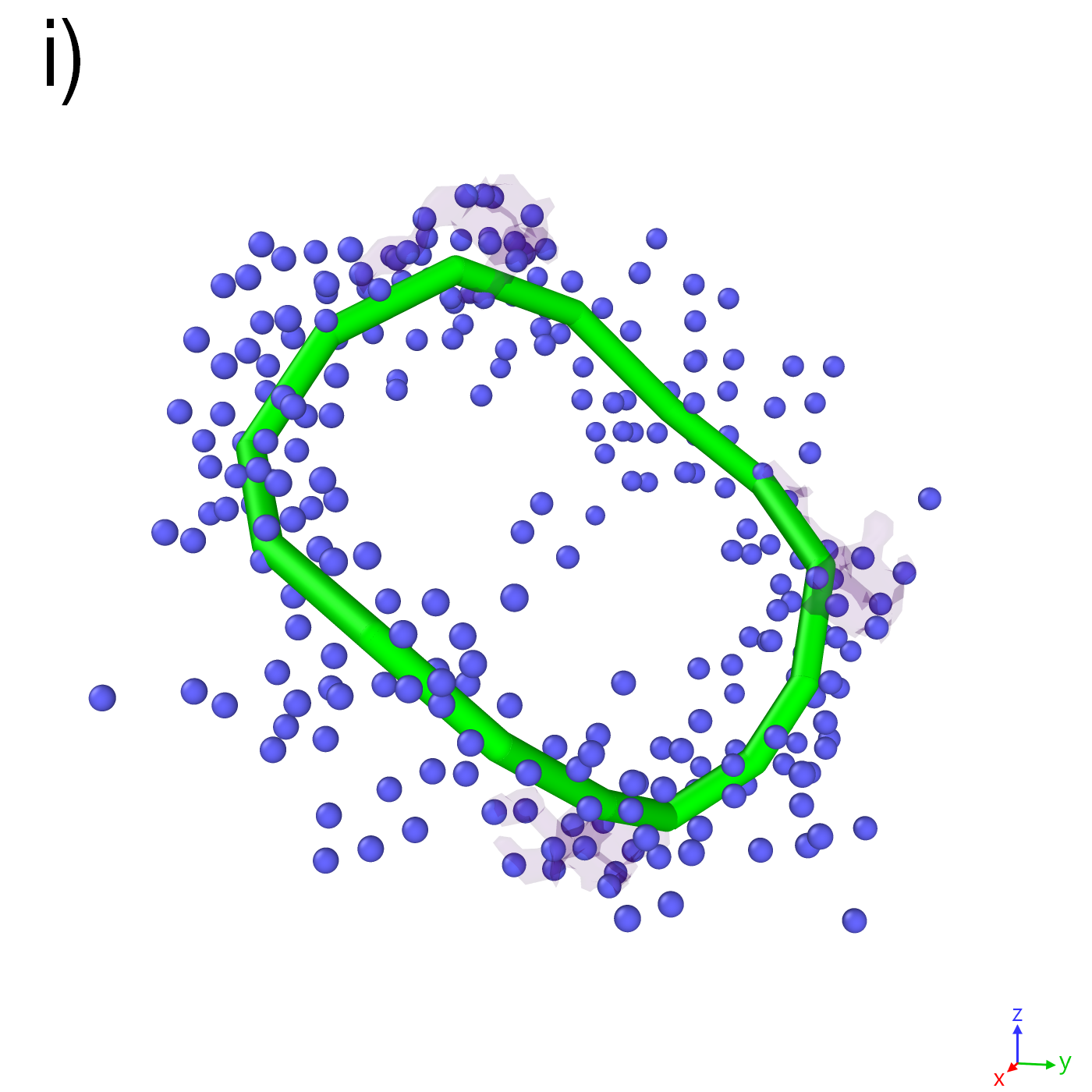}
    \includegraphics[width=0.22\linewidth]{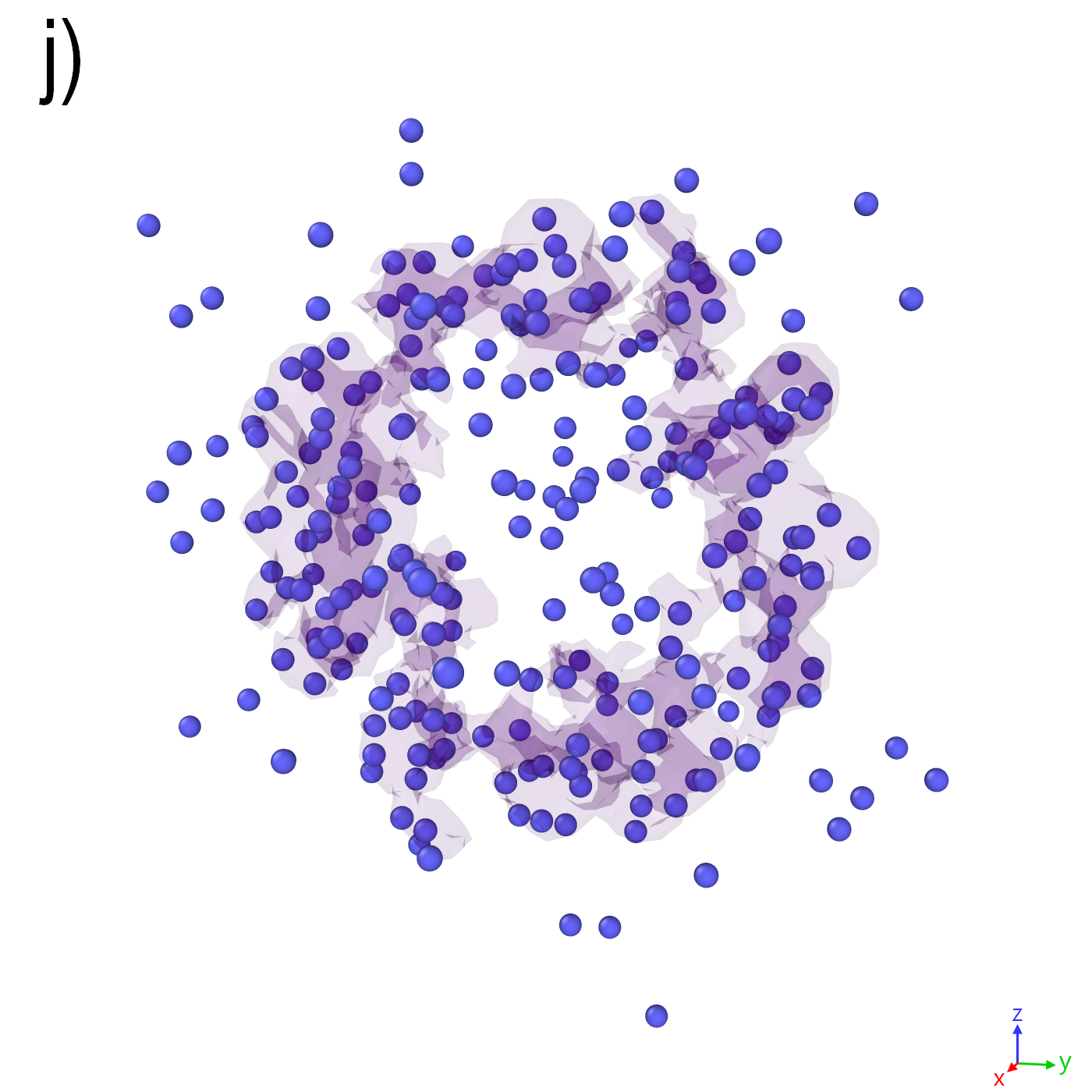}
    \includegraphics[width=0.22\linewidth]{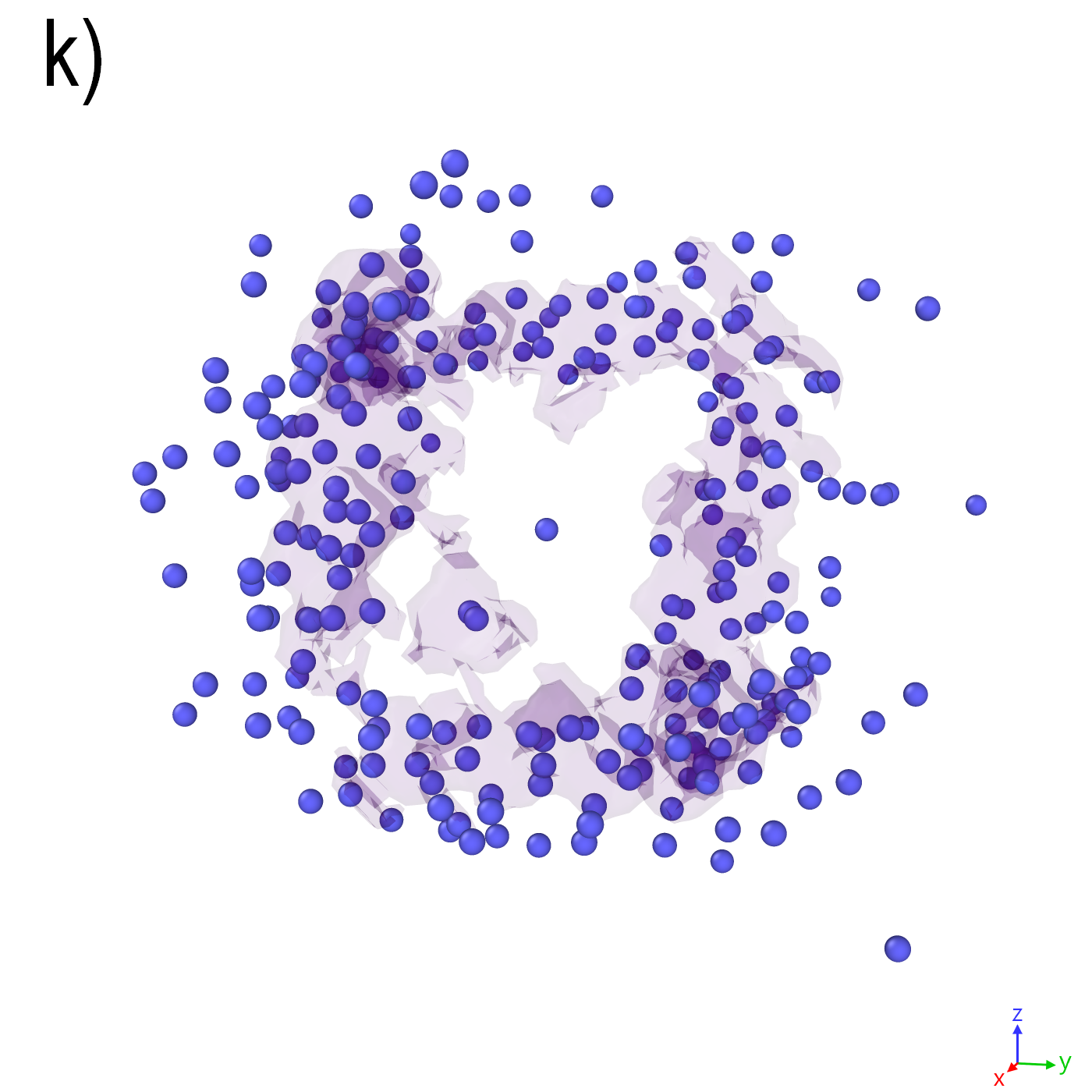}   \\
    \includegraphics[width=0.22\linewidth]{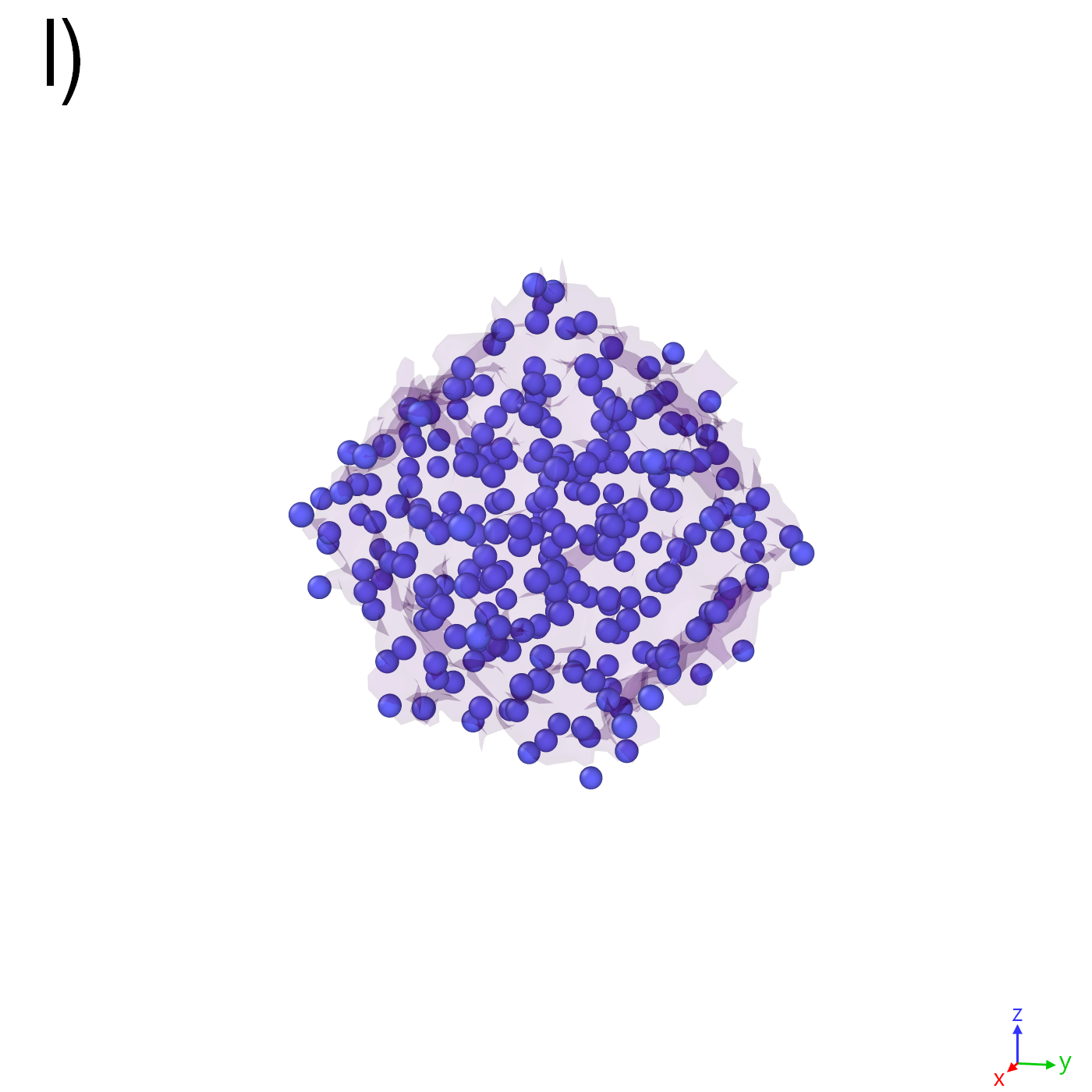}
    \includegraphics[width=0.22\linewidth]{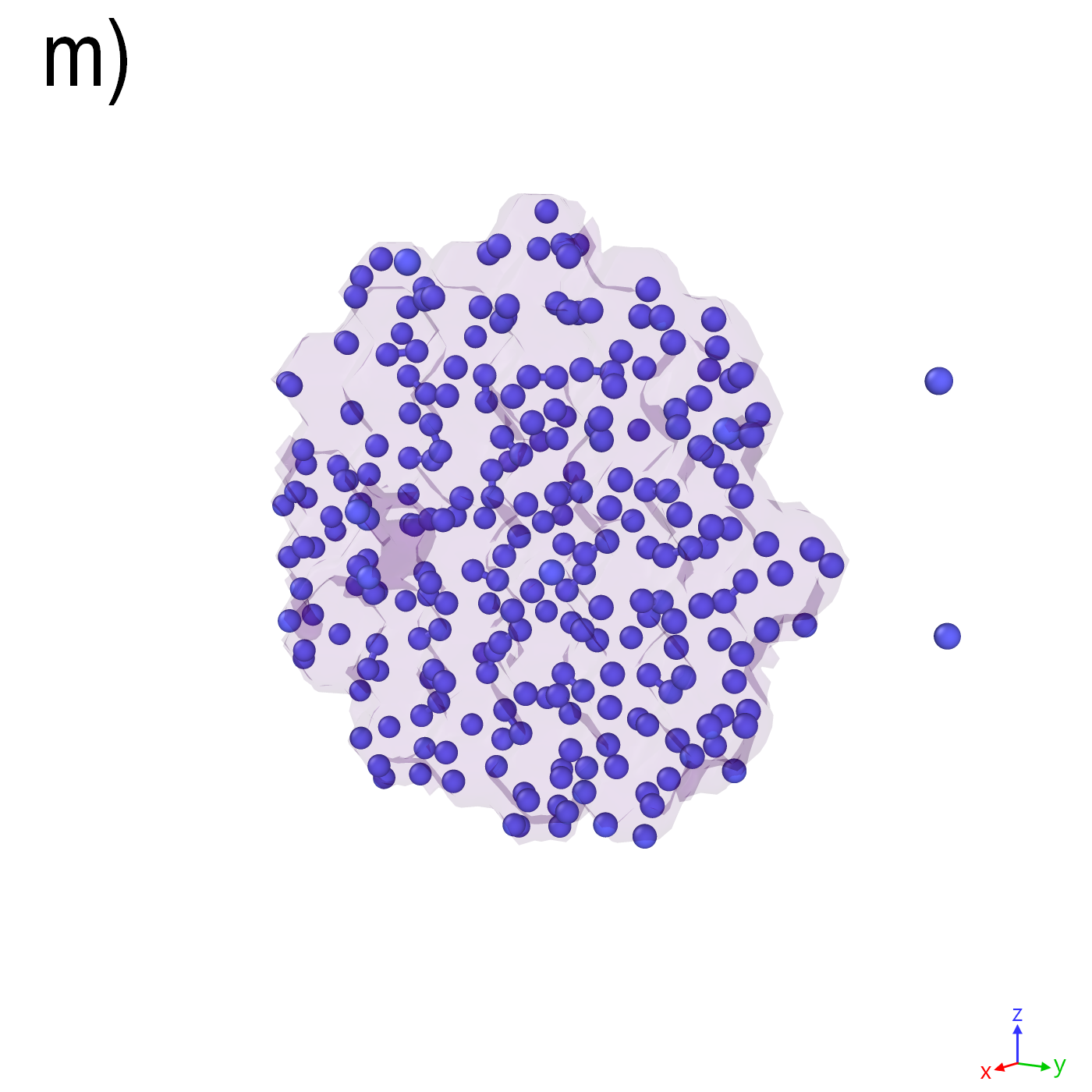}
    \includegraphics[width=0.22\linewidth]{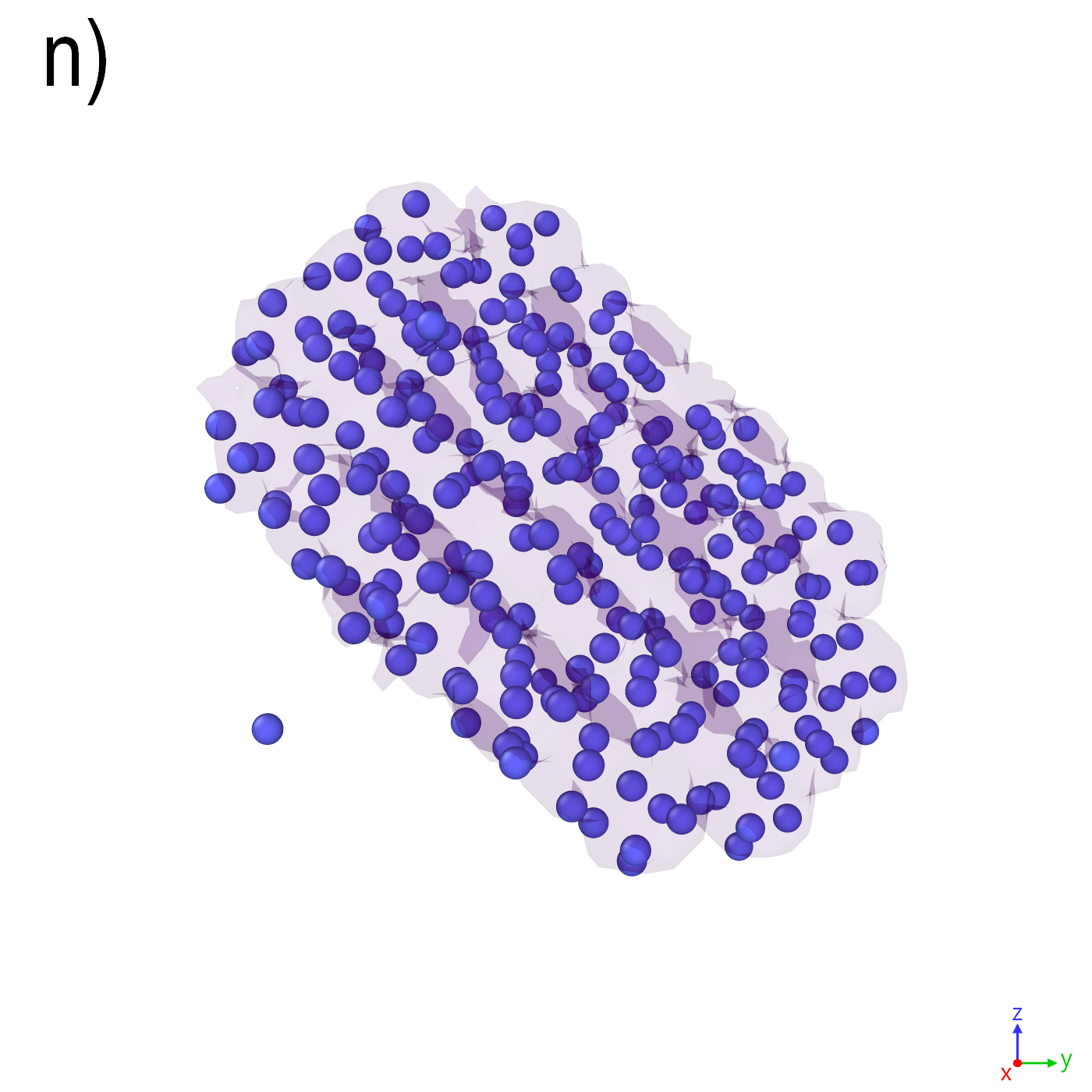}         
    \caption{
    Extended defects in tungsten decorated with multiple H atoms.
    From left-to-right , nanovoid, vacancy platelet with \hkl<100> and \half\hkl<111> Burgers vectors. 
    Each has a size of around 60 point defects, making the loops and plates $\sim 2$ nm diameter, and the void $\sim 1.5$ nm diameter.
    a) to d) - interstitial loops with \hkl<100> and \half\hkl<111> Burgers vectors, vacancy loops with \hkl<100> and \half\hkl<111> Burgers vectors, decorated with 64 H ( $n/m \sim 1$ )
    e) to g) - nanovoid, vacancy platelet with \hkl<100> and \half\hkl<111> Burgers vectors, decorated with 64 H ( $n/m \sim 1$ )
    h) to k) - as a) to d), decorated with 256 H ( $n/m \sim 4$ )
    l) to n) - as e) to g), decorated with 256 H ( $n/m \sim 4$ ).
    Pink lines show \hkl<100> edge dislocation lines, green lines show \hkl<111> edge dislocation lines.
    Purple surfaces show boundaries of cavity regions.
    Tungsten atoms not shown.
    Rendered with OVITO~\cite{Stukowski_MSMSE2009}.
    }
    \label{fig:defectRenders}
\end{figure*}

Figure \ref{fig:defectRenders} shows renderings of some of the decorated defects.
We have used the method of ref~\cite{Mason_PRM2021} to find void isosurfaces, and DXA~\cite{Stukowski_MSMSE2009,Stukowski_MSMSE2012} to find dislocation lines.
Two sets of images are shown, with filling ratios $n/m \sim 1$ and $n/m \sim 4$.
At $n/m \sim 1$, the dislocation loop defects are easily identifiable as such, with a small number of H atoms decorating the dislocation lines. 
Void spaces do arise between W atoms even at low $n/m$, but they appear as cracks, significantly thinner than a monovacancy, between W atoms under considerable tensile strain in the direction of the Burgers vector.
In the open defects ( voids and platelets ) molecular H$_2$ can be seen in the interior, with H atoms also decorating the surface.
At $n/m \sim 4$, the interstitial dislocation loops are still identifiable, but the vacancy dislocation loops have lost their cores, instead appearing closer to a thin toroidal crack between the atoms in the direction of the Burgers vector. No interstitial ejection is observed, 
and no molecular H$_2$ is observed.
The open defects show saturation at $n/m \sim 4$, and though some H$_2$ can be seen the majority of the gas atoms are on the surface of the cavities. It is notable that some H is not inside the cavities, but sits in interstitial positions beyond the defect.

\section{Conclusions}

No empirical potential is expected to perfectly reproduce all properties, especially those outside the targetted set, and this potential is no exception. We have tried to report faithfully areas where there could be concerns. 
We therefore note 
    \begin{itemize}
        \item The trivacancy migration energy in pure W is close to the monovacancy migration energy, and is not the very low value found by DFT. This is a failure of the EAM potential form to reproduce the correct sp-d hybridization and can not be easily fixed. This may have an impact in studies concerning the clustering of monovacancies.
        \item The \hkl(100) surface energy of pure W is ordered incorrectly, again as a failure of the EAM potential form to reproduce the correct sp-d hybridization. The \hkl(100) surface should be treated as least reliable in quantitative studies.
        \item The binding energy of an H atom to a surface is low, ranging from 0.3 - 0.5 eV compared to the experimental value 0.7 - 0.8 eV. While this is an improvement on previous literature potentials, it is most likely related again to the difficulty of reproducing surface electron states. This may have an impact where equilibrium between molecular H$_2$ and the surface is required.
    \end{itemize}

We have shown that the potential described here is capable of reproducing a range of DFT-calculated properties of the interactions between hydrogen isotope gases dissolved in a tungsten lattice, and point- and extended- lattice defects introduced by irradiation damage.
In particular, we have focused on ensuring that not only the binding energies, but also the relaxation volumes of the defects are well reproduced.
This gives us more confidence that the potential will show qualitatively correct trends as a function of the applied boundary conditions of stress and strain.

\section{Data Availability}

The potential in LAMMPS compatible \texttt{eam/alloy} pair style will be submitted to the NIST database on acceptance.

\section{Acknowledgements}
 
We are grateful to K. Heinola and C. Becquart for helpful discussions, and L. Yang for providing the LAMMPS potential files for their tungsten-hydrogen potential.
This work has been carried out within the framework of the EUROfusion Consortium, funded by the European Union via the Euratom Research and Training Programme (Grant Agreement No. 101052200 - EUROfusion), and by the RCUK Energy Programme, Grant No. EP/W006839/1. To obtain further information on the data and models underlying the paper please contact PublicationsManager@ukaea.uk. The views and opinions expressed herein do not necessarily reflect those of the European Commission.
 
\section{Author contributions}
D.R.M. developed the concept, fitted the potential, and generated results using the potential. D.N.-M. performed the DFT calculations and contributed to the development of the model. V.W.L. and F.G.G. performed MD testing under a range of conditions, M.N.L. generated results using empirical potentials and contributed to the development of the model.

\section{Appendix: parameterization}
\label{sec:parameterization}

In this section we present the parameterization of the potential.

The explicit form chosen for the three functions is
    \begin{eqnarray}
        V(r) &=& \left\{ \begin{array}{lcr}
           Z(r) +  P_V^{(0)}(r)                          &    \quad ,\quad  & 0 \le r \le r_V^{(1)}               \\
           Z(r) +  P_V^{(1)}(r)                          &  &     r_V^{(1)} \le r \le r_V^{(2)}       \\
                   P_V^{(2)}(r)                          &  &     r_V^{(2)} \le r \le r_V^{(3)}        \\
           0                                            &  &    r \ge r_V^{(3)} 
        \end{array} \right.         \nonumber       \\
        \phi(x) &=& \left\{ \begin{array}{lcr}
           P_{\phi}^{(0)}(r)                           &   \quad \quad \quad \quad ,\quad  &    0 \le r \le r_{\phi}^{(1)}               \\
           P_{\phi}^{(1)}(r)                           &  &     r_{\phi}^{(1)} \le r \le r_{\phi}^{(2)}       \\
           0                                        & &      r \ge r_{\phi}^{(2)}    
        \end{array} \right.         \nonumber       \\
        F(\rho) &=& \left\{ \begin{array}{lcr}
           \alpha \sqrt{\rho} + P_{F}^{(0)}(\rho)                           &   \quad ,\quad  &    0 \le \rho \le \rho^{(1)}               \\
           \alpha \sqrt{\rho} + P_{F}^{(1)}(\rho)                           &  &     \rho^{(1)} \le \rho \le \rho^{(2)}       \\
           \alpha \sqrt{\rho}                                         & &      \rho \ge \rho_{\phi}^{(2)}    
        \end{array} \right. ,             
    \end{eqnarray}  
where the functions $P_{\sigma}^{(n)}(x)$ are fifth-order polynomials which match zeroth, first, and second derivatives at the knot points $x_{\sigma}^{(n)}$. 
The full parameterization is given in table \ref{tab:parameterization}.
We also constrain by \emph{fiat} some of the values and derivatives at zero or the cutoff range, to ensure continuity and smoothness.


The functional form of $P_{\sigma}^{(n)}(x)$ can be found by substituting into the general polynomial form
    \begin{eqnarray}
        P_{\sigma}^{(n)}(z) &=& P_0 + P'_0 z + P''_0 z^2/2      \nonumber \\
            &=& - (20 P_0 - 20 P_1 + 12 P'_0 + 8 P'_1 + 3 P''_0 - P''_1 ) z^3/2    \nonumber \\
            && + (30 P_0 - 30 P_1 + 16 P'_0 + 14 P'_1 + 3 P''_0 - 2 P''_1 ) z^4/2    \nonumber \\
            && - (12 P_0 - 12 P_1 + 6 P'_0 + 6 P'_1 + P''_0 - P''_1 ) z^5/2,
    \end{eqnarray}
where $z=(x_{\sigma}^{(n)} - x)/(x_{\sigma}^{(n)} - x_{\sigma}^{(n-1)})$, and $P_{0,1},P'_{0,1},P''_{0,1}$ are the values of the function and its derivatives with respect to $z$ which are matched at either end of the interval. 
Note that in table \ref{tab:parameterization} we present the derivatives with respect to $x$ not $z$. 

To the repulsive pairwise potential we add the ZBL form between $r=0$ and $r=r_2$. This has the universal form reproduced here for completeness
    \begin{eqnarray}
        V_{ZBL}(r) = \frac{ Z_1 Z_2 }{4 \pi \varepsilon \,r}  \left( \right.& & 0.1818 e^{-3.2 x/a} + 0.5099 e^{-0.9423 x/a}    \nonumber\\
        &&  \left. + 0.2802 e^{-0.4029 x/a} + 0.02817 e^{-0.2016 x/a} \right),  \nonumber\\
    \end{eqnarray}
with $Z_{1(2)}$ being the atomic number of element $1(2)$, and $a=0.078908 \AA$ the screening length constant.

Finally, in order to increase the binding of hydrogen atoms to surfaces, we further add a long range attractive interaction to the $V_{WH}$ potential, of the form $V^+(r) = V_0$, $r<r_3$, tapering to zero at $r=r_4$.

\begin{table*}[htb!]
    \centering
    \begin{tabular}{lr|lr|lr|lr}    
    $F_W(\rho)$         \\
    \hline
    $\alpha$         &   -10.39248715    \\
    $\rho_0$         &   0               &   $F(\rho_0)$    &   0               &  $F'(\rho_0)$    &   2.06957725  &  $F''(\rho_0)$    &  -5.11678124     \\
    $\rho_1$         &   0.66059743      &   $F(\rho_1)$    &   0.61968010      &  $F'(\rho_1)$    &  -1.07962770  &  $F''(\rho_1)$    &  0.95886630      \\
    $\rho_2$         &   1.05226457      &   $F(\rho_2)$    &   0               &  $F'(\rho_2)$    &   0           &  $F''(\rho_2)$    &   0              \\
    \hline
    $\phi_W(r)$         \\
    \hline
    $r_0$         &   0               &   $\phi(r_0)$    &   0.51304656      &  $\phi'(r_0)$    &  -0.26614197  &  $\phi''(r_0)$    &  0.06274133     \\
    $r_1$         &   3.27332980      &   $\phi(r_1)$    &   0.04521507      &  $\phi'(r_1)$    &  -0.07499284  &  $\phi''(r_1)$    &  0.06432726      \\
    $r_2$         &   4.64995591      &   $\phi(r_2)$    &   0               &  $\phi'(r_2)$    &   0           &  $\phi''(r_2)$    &   0              \\
    \hline
    $F_H(\rho)$         \\
    \hline
    $\alpha$         &   -2.094997341   \\
    $\rho_0$         &   0               &   $F(\rho_0)$    &   0               &  $F'(\rho_0)$    &  -0.999828063  &  $F''(\rho_0)$    &  13.29712796     \\
    $\rho_1$         &    2.206458605     &   $F(\rho_1)$    &   3.0265026      &  $F'(\rho_1)$    &  2.597429616   &  $F''(\rho_1)$    &  6.826075643
      \\
    $\rho_2$         &  232.4702459     &   $F(\rho_2)$    &   0               &  $F'(\rho_2)$    &   0           &  $F''(\rho_2)$    &   0              \\
    \hline
    $\phi_H(r)$         \\
    \hline
    $r_0$         &   0               &   $\phi(r_0)$    &   3.15074682      &  $\phi'(r_0)$    &   0.08150179  &  $\phi''(r_0)$    &  0.02370387      \\
    $r_1$         &   1.73956971      &   $\phi(r_1)$    &   0.21459106      &  $\phi'(r_1)$    &  -0.58558991  &  $\phi''(r_1)$    &  1.40127598      \\
    $r_2$         &   3.05953885      &   $\phi(r_2)$    &   0               &  $\phi'(r_2)$    &   0           &  $\phi''(r_2)$    &   0              \\
    \hline
    $V_{WW}(r)$         \\
    \hline
    $r_0$         &   0               &   $V(r_0)$    &   0               &  $V'(r_0)$    &   0           &  $V''(r_0)$    &   0              \\
    $r_1$         &   2.75682338      &   $V(r_1)$    &  -0.36045809      &  $V'(r_1)$    &  0.78813164   &  $V''(r_1)$    &  -1.38844445     \\
    $r_2$         &   3.09550508      &   $V(r_2)$    &  -0.16134421      &  $V'(r_2)$    &  0.23211407   &  $V''(r_2)$    &   3.09284796     \\
    $r_3$         &   4.44746515      &   $V(r_3)$    &   0               &  $V'(r_3)$    &   0           &  $V''(r_3)$    &   0              \\
    \hline    
    $V_{WH}(r)$         \\
    \hline
    $r_0$         &   0               &   $V(r_0)$    &   0               &  $V'(r_0)$    &   0           &  $V''(r_0)$    &   0              \\
    $r_1$         &   1.49377362      &   $V(r_1)$    &  2.51539938       &  $V'(r_1)$    &  -6.19359099  &  $V''(r_1)$    &  11.06601825     \\
    $r_2$         &   0.01865712      &   $V(r_2)$    &  0.01865712       &  $V'(r_2)$    &  -0.26353573  &  $V''(r_2)$    &   0.16942984     \\
    $r_3$         &   3.92905810      &   $V(r_3)$    &   0               &  $V'(r_3)$    &   0           &  $V''(r_3)$    &   0              \\
    $r_4$         &   4.84477478      &   $\Delta V$  &  -0.10056716      &               &               &                &   \\
    \hline    
    $V_{HH}(r)$         \\
    \hline
    $r_0$         &   0               &   $V(r_0)$    &   0               &  $V'(r_0)$    &   0           &  $V''(r_0)$    &   0              \\
    $r_1$         &   0.29932596      &   $V(r_1)$    &  -0.00277562      &  $V'(r_1)$    &  -0.01103995  &  $V''(r_1)$    &  504.9017561     \\
    $r_2$         &   0.73972733      &   $V(r_2)$    &  -4.65387843      &  $V'(r_2)$    &  14.21149452  &  $V''(r_2)$    &  -74.74055820     \\
    $r_3$         &   1.41269052      &   $V(r_3)$    &   0               &  $V'(r_3)$    &   0           &  $V''(r_3)$    &   0              \\
    \hline    
    \end{tabular}
    \caption{Full parameterization of the potential, defined as function values and derivatives at knot points. Lengths ($x$) in \AA, energies ($V,F$) in eV.}
    \label{tab:parameterization}
\end{table*}


\section*{References}
\bibliographystyle{unsrt}
\bibliography{mnl6_arxiv}

\begin{thebibliography}{10}

\bibitem{You2016}
J.~H. You, E.~Visca, Ch. Bachmann, T.~Barrett, F.~Crescenzi, M.~Fursdon,
  H.~Greuner, D.~Guilhem, P.~Languille, M.~Li, S.~McIntosh, A.~V. Müller,
  J.~Reiser, M.~Richou, and M.~Rieth.
\newblock European {DEMO} divertor target: Operational requirements and
  material-design interface.
\newblock {\em Nuclear Materials and Energy}, 9:171--176, 2016.

\bibitem{Gilbert_FST2014}
M.~R. Gilbert, S.~Zheng, R.~Kemp, L.~W. Packer, S.~L. Dudarev, and J.-Ch.
  Sublet.
\newblock Comparative assessment of material performance in {DEMO} fusion
  reactors.
\newblock {\em Fusion Science and Technology}, 66(1):9--17, 2014.

\bibitem{Keys_PhysRev_1968}
L.~K. Keys, J.~P. Smith, and J.~Moteff.
\newblock High-temperature recovery of tungsten after neutron irradiation.
\newblock {\em Phys. Rev.}, 176:851--856, 1968.

\bibitem{Hirai_NME2016}
T.~Hirai, S.~Panayotis, V.~Barabash, C.~Amzallag, F.~Escourbiac, A.~Durocher,
  M.~Merola, J.~Linke, Th. Loewenhoff, G.~Pintsuk, M.~Wirtz, and
  I.~Uytdenhouwen.
\newblock Use of tungsten material for the iter divertor.
\newblock {\em Nuclear Materials and Energy}, 9:616--622, 2016.

\bibitem{Li_NME2018}
Muyuan Li and Jeong-Ha You.
\newblock Structural impact of creep in tungsten monoblock divertor target at
  20 mw/m2.
\newblock {\em Nuclear Materials and Energy}, 14:1--7, 2018.

\bibitem{You_JNM2021}
J.H. You, E.~Visca, T.~Barrett, B.~Böswirth, F.~Crescenzi, F.~Domptail,
  G.~Dose, M.~Fursdon, F.~Gallay, H.~Greuner, K.~Hunger, A.~Lukenskas, A.v.
  Müller, M.~Richou, S.~Roccella, C.~Vorpahl, and K.~Zhang.
\newblock High-heat-flux technologies for the european demo divertor targets:
  State-of-the-art and a review of the latest testing campaign.
\newblock {\em Journal of Nuclear Materials}, 544:152670, 2021.

\bibitem{Ogorodnikova_JNM2015}
O.V. Ogorodnikova and V.~Gann.
\newblock Simulation of neutron-induced damage in tungsten by irradiation with
  energetic self-ions.
\newblock {\em Journal of Nuclear Materials}, 460:60--71, 2015.

\bibitem{Ogorodnikova_JAP2015}
O.~V. Ogorodnikova.
\newblock Fundamental aspects of deuterium retention in tungsten at high flux
  plasma exposure.
\newblock {\em J. Appl. Phys}, 118:074902, 2015.

\bibitem{DeBacker_PhysScr2017}
A.~De~Backer, D~R Mason, C~Domain, D~Nguyen-Manh, M-C Marinica, L~Ventelon, C~S
  Becquart, and S~L Dudarev.
\newblock Hydrogen accumulation around dislocation loops and edge dislocations:
  from atomistic to mesoscopic scales in {BCC} tungsten.
\newblock {\em Physica Scripta}, T170:014073, nov 2017.

\bibitem{DeBacker_NucFus2018}
A.~De~Backer, D.~R. Mason, C.~Domain, D.~Nguyen-Manh, M.~C. Marinica,
  L.~Ventelon, C.~S. Becquart, and S.~R. Dudarev.
\newblock Multiscale modelling of the interaction of hydrogen with interstitial
  defects and dislocations in bcc tungsten.
\newblock {\em Nuclear Fusion}, 58:016006, 2018.

\bibitem{Wang_JNM2021}
Jing Wang, Yuji Hatano, Tatsuya Hinoki, Vladimir~Kh. Alimov, Alexander~V.
  Spitsyn, Nikolay~P. Bobyr, Sosuke Kondo, Takeshi Toyama, Heun~Tae Lee, Yoshio
  Ueda, and Thomas Schwarz-Selinger.
\newblock Deuterium retention in {W} and binary {W} alloys irradiated with high
  energy fe ions.
\newblock {\em Journal of Nuclear Materials}, 545:152749, 2021.

\bibitem{Hollingsworth_NF2019}
A.~Hollingsworth, M.Yu. Lavrentiev, R.~Watkins, A.C. Davies, S.~Davies,
  R.~Smith, D.R. Mason, A.~Baron-Wiechec, Z.~Kollo, J.~Hess, I.~Jepu,
  J.~Likonen, K.~Heinola, K.~Mizohata, E.~Meslin, M.-F. Barthe, A.~Widdowson,
  I.S. Grech, K.~Abraham, E.~Pender, A.~McShee, Y.~Martynova, M.~Freisinger,
  and A.~De~Backer.
\newblock Comparative study of deuterium retention in irradiated eurofer and
  {Fe{\textendash}Cr} from a new ion implantation materials facility.
\newblock {\em Nuclear Fusion}, 60(1):016024, nov 2019.

\bibitem{Mason_PRL2020}
Daniel~R. Mason, Suchandrima Das, Peter~M. Derlet, Sergei~L. Dudarev, Andrew~J.
  London, Hongbing Yu, Nicholas~W. Phillips, David Yang, Kenichiro Mizohata,
  Ruqing Xu, and Felix Hofmann.
\newblock Observation of transient and asymptotic driven structural states of
  tungsten exposed to radiation.
\newblock {\em Phys. Rev. Lett.}, 125:225503, Nov 2020.

\bibitem{Kato_NucFus2015}
D.~Kato, H.~Iwakiri, Y.~Watanabe, K.~Morishita, and T.~Muroga.
\newblock Super-saturated hydrogen effects on radiation damages in tungsten
  under the high-flux divertor plasma irradiation.
\newblock {\em Nuclear Fusion}, 55(8):083019, jul 2015.

\bibitem{SchwarzSelinger_NME2018}
T.~Schwarz-Selinger, J.~Bauer, S.~Elgeti, and S.~Markelj.
\newblock Influence of the presence of deuterium on displacement damage in
  tungsten.
\newblock {\em Nuclear Materials and Energy}, 17:228--234, 2018.

\bibitem{Liu_AIPA2013}
Y-N. Liu, T.~Ahlgren, L.~Bukonte, K.~Nordlund, X.~Shu, Y.~Yu, X-C. Li, and G-H.
  Lu.
\newblock Mechanism of vacancy formation induced by hydrogen in tungsten.
\newblock {\em AIP Advances}, page 122111, 2013.

\bibitem{Qin_JNM2015}
Shi-Yao Qin, Shuo Jin, Lu~Sun, Hong-Bo Zhou, Ying Zhang, and Guang-Hong Lu.
\newblock Hydrogen assisted vacancy formation in tungsten: A first-principles
  investigation.
\newblock {\em Journal of Nuclear Materials}, 465:135--141, 2015.

\bibitem{Zayachuk_NucFus2013}
Y.~Zayachuk, M.H.J. 't~Hoen, P.A.~Zeijlmans van Emmichoven, D.~Terentyev,
  I.~Uytdenhouwen, and G.~van Oost.
\newblock Surface modification of tungsten and tungsten{\textendash}tantalum
  alloys exposed to high-flux deuterium plasma and its impact on deuterium
  retention.
\newblock {\em Nuclear Fusion}, 53(1):013013, jan 2013.

\bibitem{Hodille_PhysRevMat2018}
E.~A. Hodille, N.~Fernandez, Z.~A. Piazza, M.~Ajmalghan, and Y.~Ferro.
\newblock Hydrogen supersaturated layers in {H/D} plasma-loaded tungsten: A
  global model based on thermodynamics, kinetics and density functional theory
  data.
\newblock {\em Physical Review Materials}, 2(9):093802, 2018.

\bibitem{Nguyen_NIMB2015}
D.~Nguyen-Manh and S.~L. Dudarev.
\newblock Trapping of he clusters by inert-gas impurities in tungsten:
  First-principles predictions and experimental validation.
\newblock {\em Nucl. Instr. Meth. B}, 352:86--91, 2015.

\bibitem{Hofmann_Acta2015}
F.~Hofmann, D.~Nguyen-Manh, M.R. Gilbert, C.E. Beck, J.~K. Eliason, A.A.
  Maznev, W.~Liu, D.~E.~J. Armstrong, K.~A. Nelson, and S.~L. Dudarev.
\newblock Lattice swelling and modulus change in a helium-implanted tungsten
  alloy: X-ray micro-diffraction, surface acoustic wave measurements, and
  multiscale modelling.
\newblock {\em Acta Materialia}, 89:352--363, 2015.

\bibitem{Mason_JPCM2017}
D~R Mason, D~Nguyen-Manh, and C~S Becquart.
\newblock An empirical potential for simulating vacancy clusters in tungsten.
\newblock {\em Journal of Physics: Condensed Matter}, 29(50):505501, 2017.

\bibitem{Mason_2019}
D.R. Mason, D.~Nguyen-Manh, M.-C. Marinica, R.~Alexander, and S.L. Dudarev.
\newblock Relaxation volumes of microscopic and mesoscopic irradiation-induced
  defects in tungsten.
\newblock {\em submitted}, 2019.

\bibitem{Wrobel_COMMAT2021}
J.~S. Wrobel, M.~R. Zembla, D.~Nguyen-Manh, P.~Olsson, L.~Messina, C.~Domain,
  T.~Wejrzanowski, and S.~L. Dudarev.
\newblock Elastic dipole tensors and relaxation volumes of point defects in
  concentrated random magnetic fe-cr alloys.
\newblock {\em Computational Materials Science}, 194:110435, 2021.

\bibitem{Wang_JPCM2017}
Li-Fang Wang, Xiaolin Shu, Guang-Hong Lu, and Fei Gao.
\newblock Embedded-atom method potential for modeling hydrogen and
  hydrogen-defect interaction in tungsten.
\newblock {\em Journal of Physics: Condensed Matter}, 29(43):435401, sep 2017.

\bibitem{Muzyk_PRB2011}
M.~Muzyk, D.~Nguyen-Manh, K.~J. Kurzyd\l{}owski, N.~L. Baluc, and S.~L.
  Dudarev.
\newblock Phase stability, point defects, and elastic properties of w-v and
  w-ta alloys.
\newblock {\em Phys. Rev. B}, 84:104115, Sep 2011.

\bibitem{Mason_PRM2021}
Daniel~R. Mason, Fredric Granberg, Max Boleininger, Thomas Schwarz-Selinger,
  Kai Nordlund, and Sergei~L. Dudarev.
\newblock Parameter-free quantitative simulation of high-dose microstructure
  and hydrogen retention in ion-irradiated tungsten.
\newblock {\em Phys. Rev. Mater.}, 5:095403, Sep 2021.

\bibitem{Simmonds_JNM2017}
M.J. Simmonds, Y.Q. Wang, J.L. Barton, M.J. Baldwin, J.H. Yu, R.P. Doerner, and
  G.R. Tynan.
\newblock Reduced deuterium retention in simultaneously damaged and annealed
  tungsten.
\newblock {\em Journal of Nuclear Materials}, 494:67--71, 2017.

\bibitem{Simmonds_NF2022}
M.J. Simmonds, T.~Schwarz-Selinger, M.I. Patino, M.J. Baldwin, R.P. Doerner,
  and G.R. Tynan.
\newblock Reduced defect recovery in self-ion damaged w due to simultaneous
  deuterium exposure during annealing.
\newblock {\em Nuclear Fusion}, 62(3):036012, jan 2022.

\bibitem{Daw_PRB1984}
Murray~S. Daw and M.~I. Baskes.
\newblock Embedded-atom method: Derivation and application to impurities,
  surfaces, and other defects in metals.
\newblock {\em Phys. Rev. B}, 29:6443--6453, Jun 1984.

\bibitem{LAMMPS}
Steve Plimpton.
\newblock Fast parallel algorithms for short-range molecular dynamics.
\newblock {\em Journal of Computational Physics}, 117(1):1 -- 19, 1995.

\bibitem{Finnis_PMA1984}
M.~W. Finnis and J.~E. Sinclair.
\newblock A simple empirical n-body potential for transition metals.
\newblock {\em Philosophical Magazine A}, 50(1):45--55, 1984.

\bibitem{Ackland_PMA1987}
G.J. Ackland and R.~Thetford.
\newblock An improved n-body semi-empirical model for body-centred cubic
  transition metals.
\newblock {\em Philosophical Magazine A}, 56:15--30, 1987.

\bibitem{Bjorkas_NIMB2009}
C.~Bj{\"o}rkas, K.~Nordlund, and S.~L. Dudarev.
\newblock Modelling radiation effects using the ab-initio based tungsten and
  vanadium potentials.
\newblock {\em Nucl. Instr. Meth. B}, 267:3204--3208, 2009.

\bibitem{Becquart_JNM2021}
Charlotte~S. Becquart, Andrée {De Backer}, Pär Olsson, and Christophe Domain.
\newblock Modelling the primary damage in fe and w: Influence of the short
  range interactions on the cascade properties: Part 1 – energy transfer.
\newblock {\em Journal of Nuclear Materials}, 547:152816, 2021.

\bibitem{Juslin_JNM2013}
N.~Juslin and B.D. Wirth.
\newblock Interatomic potentials for simulation of he bubble formation in w.
\newblock {\em Journal of Nuclear Materials}, 432(1):61 -- 66, 2013.

\bibitem{SRIM}
J.F. Ziegler, J.P. Biersack, and U.~Littmark.
\newblock {\em The stopping and range of ions in solids}.
\newblock Pergamon, 1982.

\bibitem{Ackland_JPhysF1988}
G~J Ackland, M~W Finnis, and V~Vitek.
\newblock Validity of the second moment tight-binding model.
\newblock {\em Journal of Physics F: Metal Physics}, 18(8):L153, aug 1988.

\bibitem{Marinica_JPCM2013}
M.-C. Marinica, L.~Ventelon, M.~R. Gilbert, L.~Proville, S.~L. Dudarev,
  J.~Marian, G.~Bencteux, and F.~Willaime.
\newblock Interatomic potentials for modelling radiation defects and
  dislocations in tungsten.
\newblock {\em Journal of Physics: Condensed Matter}, 25(39):395502, 2013.

\bibitem{Bonny_JPCM2014}
G.~Bonny, P.~Grigorev, and D.~Terentyev.
\newblock On the binding of nanometric hydrogen-helium clusters in tungsten.
\newblock {\em Journal of Physics: Condensed Matter}, 26(48):485001, 2014.

\bibitem{Heinola_NF2018}
K.~Heinola, F.~Djurabekova, and T.~Ahlgren.
\newblock On the stability and mobility of di-vacancies in tungsten.
\newblock {\em Nuclear Fusion}, 58(2):026004, dec 2017.

\bibitem{Kong_Acta2015}
K.~Heinola, F.~Djurabekova, and T.~Ahlgren.
\newblock First-principles calculations of hydrogen solution and diffusion in tungsten: Temperature and defect-trapping effects.
\newblock {\em Acta Materialia}, 84:426-435, 2015.

\bibitem{Mason_JAP2019}
Daniel~R. Mason, Duc Nguyen-Manh, Mihai-Cosmin Marinica, Rebecca Alexander,
  Andrea~E. Sand, and Sergei~L. Dudarev.
\newblock Relaxation volumes of microscopic and mesoscopic irradiation-induced
  defects in tungsten.
\newblock {\em Journal of Applied Physics}, 126(7):075112, 2019.

\bibitem{Featherston_PR1963}
F.~H. Featherston and J.~R. Neighbours.
\newblock Elastic constants of tantalum, tungsten, and molybdenum.
\newblock {\em Phys. Rev.}, 130:1324--1333, May 1963.

\bibitem{Varvenne_PRB2013}
C\'eline Varvenne, Fabien Bruneval, Mihai-Cosmin Marinica, and Emmanuel Clouet.
\newblock Point defect modeling in materials: Coupling ab initio and elasticity
  approaches.
\newblock {\em Phys. Rev. B}, 88:134102, Oct 2013.

\bibitem{Ma_PRM2020}
Pui-Wai Ma, D.~R. Mason, and S.~L. Dudarev.
\newblock Multiscale analysis of dislocation loops and voids in tungsten.
\newblock {\em Phys. Rev. Materials}, 4:103609, Oct 2020.

\bibitem{Methfessel1992}
M.~Methfessel, D.~Hennig, and M.~Scheffler.
\newblock Trends of the surface relaxations, surface energies, and work
  functions of the 4d transition metals.
\newblock {\em Phys. Rev. B}, 46:4816, Aug 1992.

\bibitem{Tyson_SurfSci1977}
W.R. Tyson and W.A. Miller.
\newblock Surface free energies of solid metals: Estimation from liquid surface
  tension measurements.
\newblock {\em Surface Science}, 62(1):267--276, 1977.

\bibitem{Alexander_PRB2016}
R.~Alexander, M.-C. Marinica, L.~Proville, F.~Willaime, K.~Arakawa, M.~R.
  Gilbert, and S.~L. Dudarev.
\newblock Ab initio scaling laws for the formation energy of nanosized
  interstitial defect clusters in iron, tungsten, and vanadium.
\newblock {\em Phys. Rev. B}, 94:024103, Jul 2016.

\bibitem{Byggmastar_PRB2019}
J.~Byggm\"astar, A.~Hamedani, K.~Nordlund, and F.~Djurabekova.
\newblock Machine-learning interatomic potential for radiation damage and
  defects in tungsten.
\newblock {\em Phys. Rev. B}, 100:144105, Oct 2019.

\bibitem{Derlet_PRB2007}
P.~M. Derlet, D.~Nguyen-Manh, and S.~L. Dudarev.
\newblock Multiscale modeling of crowdion and vacancy defects in
  body-centered-cubic transition metals.
\newblock {\em Phys. Rev. B}, 76:054107, Aug 2007.

\bibitem{Gra20}
F.~Granberg, J.~Byggmästar, and K.~Nordlund.
\newblock Defect accumulation and evolution during prolonged irradiation of
  {F}e and {F}e{C}r alloys.
\newblock {\em J. Nucl. Mater.}, 528:151843, 2020.

\bibitem{Gra21}
F.~Granberg, J.~Byggm\"astar, and K.~Nordlund.
\newblock Molecular dynamics simulations of high-dose damage production and
  defect evolution in tungsten.
\newblock {\em Journal of Nuclear Materials}, 556:153158, 2021.

\bibitem{Derlet_PRM2020}
P.~M. Derlet and S.~L. Dudarev.
\newblock Microscopic structure of a heavily irradiated material.
\newblock {\em Phys. Rev. Materials}, 4:023605, Feb 2020.

\bibitem{Yi_EPL2015}
X.~Yi, A.~E. Sand, D.~R. Mason, M.~A. Kirk, S.~G. Roberts, K.~Nordlund, and
  S.~L. Dudarev.
\newblock Direct observation of size scaling and elastic interaction between
  nano-scale defects in collision cascades.
\newblock {\em EPL (Europhysics Letters)}, 110(3):36001, 2015.

\bibitem{Fernandez_ActaMater2015}
N.~Fernandez, Y.~Ferro, and D.~Kato.
\newblock Hydrogen diffusion and vacancies formation in tungsten: Density
  functional theory calculations and statistical models.
\newblock {\em Acta Mater}, 94:307--318, 2015.

\bibitem{Liu_JNM2009}
Y.-L. Liu, Y.~Zhang, G.N. Luo, and G.-H. Lu.
\newblock Structure, stability and diffusion of hydrogen in tungsten: A
  first-principles study.
\newblock {\em J. Nucl. Mater.}, 390-391:1032--1034, 2009.

\bibitem{Heinola_PRB2010b}
K.~Heinola, T.~Ahlgren, K.~Nordlund, and J.~Keinonen.
\newblock Hydrogen interaction with point defects in tungsten.
\newblock {\em Phys. Rev. B}, 82:094102, Sep 2010.

\bibitem{Johnson_JMR2010}
D.F. Johnson and E.A Carter.
\newblock Hydrogen in tungsten: Absorption, diffusion, vacancy trapping, and
  decohesion.
\newblock {\em J. Mater. Res}, 25:315--327, 2010.

\bibitem{PhysRevB.47.558}
G.~Kresse and J.~Hafner.
\newblock Ab initio molecular dynamics for liquid metals.
\newblock {\em Phys. Rev. B}, 47:558--561, Jan 1993.

\bibitem{PhysRevB.54.11169}
G.~Kresse and J.~Furthm{\"u}ller.
\newblock Efficient iterative schemes for ab initio total-energy calculations
  using a plane-wave basis set.
\newblock {\em Phys. Rev. B}, 54:11169--11186, Oct 1996.

\bibitem{KRESSE199615}
G.~Kresse and J.~Furthm{\"u}ller.
\newblock Efficiency of ab-initio total energy calculations for metals and
  semiconductors using a plane-wave basis set.
\newblock {\em Computational Materials Science}, 6(1):15 -- 50, 1996.

\bibitem{NguyenManh_PRB2006}
D.~Nguyen-Manh, A.~P. Horsfield, and S.L. Dudarev.
\newblock Self-interstitial atom defects in bcc transition metals:
  Group-specific trends.
\newblock {\em Physical Review B}, 73:020101, 2006.

\bibitem{Nguyen_JMS2012}
D.~Nguyen-Manh and S.~L. Dudarev.
\newblock First-principles models for phase stability and radiation defects in
  structural materials for future fusion power-plant applications.
\newblock {\em Journal of Materials Science}, 47:7385--7389, 2012.

\bibitem{Perdew_PRL1996}
J.~P. Perdew, K.~Burke, and M.~Ernzerhof.
\newblock Generalized gradient approximation made simple.
\newblock {\em Phys. Rev. Lett.}, 77:3865--3868, Oct 1996.

\bibitem{Heinola_PRB2010}
K.~Heinola and T.~Ahlgren.
\newblock First-principles study of {H} on the reconstructed {W(100)} surface.
\newblock {\em Phys. Rev. B}, 81:073409, Feb 2010.

\bibitem{Ohsawa_JNM2007}
Kazuhito Ohsawa and Eiichi Kuramoto.
\newblock Thermally activated transport of a dislocation loop within an elastic
  model.
\newblock {\em Journal of Nuclear Materials}, 367-370:327 -- 331, 2007.
\newblock Proceedings of the Twelfth International Conference on Fusion Reactor
  Materials (ICFRM-12).

\bibitem{Hou_NatMat2019}
J.~Hou, X.-S. Kong, X.~Wu, J.~Song, and C.S. Liu.
\newblock Predictive model of hydrogen trapping and bubbling in nanovoids in
  bcc metals.
\newblock {\em Nature Mater}, 18:833--839, 2019.

\bibitem{Jenkins2001}
M.L. Jenkins and M.A. Kirk.
\newblock {\em Characterization of Radiation Damage by Transmission Electron
  Microscopy}.
\newblock Series in Microscopy in Materials Science. IOP, Bristol, 2001.

\bibitem{Dudarev_Acta2017}
S.L. Dudarev and A.P. Sutton.
\newblock Elastic interactions between nano-scale defects in irradiated
  materials.
\newblock {\em Acta Materialia}, 125:425 -- 430, 2017.

\bibitem{DeBacker_NucFus2017}
A.~De~Backer, D.R. Mason, C.~Domain, D.~Nguyen-Manh, M.-C. Marinica,
  L.~Ventelon, C.S. Becquart, and S.L. Dudarev.
\newblock Multiscale modelling of the interaction of hydrogen with interstitial
  defects and dislocations in {BCC} tungsten.
\newblock {\em Nuclear Fusion}, 58(1):016006, nov 2017.

\bibitem{Gilbert_JPCM2008}
M~R Gilbert, S~L Dudarev, P~M Derlet, and D~G Pettifor.
\newblock Structure and metastability of mesoscopic vacancy and interstitial
  loop defects in iron and tungsten.
\newblock {\em Journal of Physics: Condensed Matter}, 20(34):345214, aug 2008.

\bibitem{Stukowski_MSMSE2009}
Alexander Stukowski.
\newblock Visualization and analysis of atomistic simulation data with
  {OVITO}{\textendash}the open visualization tool.
\newblock {\em Modelling and Simulation in Materials Science and Engineering},
  18(1):015012, dec 2009.

\bibitem{Stukowski_MSMSE2012}
Alexander Stukowski.
\newblock Structure identification methods for atomistic simulations of
  crystalline materials.
\newblock {\em Modelling and Simulation in Materials Science and Engineering},
  20(4):045021, may 2012.

\end{thebibliography}

\end{document}